\documentclass[12pt]{article}
\usepackage{textcomp}
\usepackage{alexchet}
\usepackage{amssymb,amsmath,amsfonts,mathrsfs,braket,mathtools}
\usepackage{pgfplots}
\usepackage[utf8]{inputenc}
\usepackage{graphicx,longtable,afterpage}
\usepackage{xcolor,hyperref}
\usepackage{xurl}

\usepackage{subcaption}
\usepackage{microtype}
\usepackage{booktabs}
\usepackage{multirow}
\usepackage{hyphenat}
\usepackage[capitalize]{cleveref}
\usepackage[normalem]{ulem}
\usepackage{tikz}
\usepackage{floatrow}

\newcommand{\sbl}{\texttt{scalar\_blocks }}

\newcommand{\gbl}{\texttt{GoBlocks }}

\usetikzlibrary{shapes.geometric, arrows, arrows.meta, backgrounds, fit, positioning,calc}

\tikzstyle{startstop} = [ellipse, draw, fill=green!20, text width=6cm, minimum height=1cm, align=center]
\tikzstyle{process} = [rectangle, draw, fill=blue!10, text width=6cm, minimum height=1cm, align=center]
\tikzstyle{decision} = [diamond, draw, fill=yellow!20, text width=4cm, minimum height=1cm, align=center, aspect=2]
\tikzstyle{parallel} = [rectangle, draw, fill=cyan!10, text width=6cm, minimum height=1cm, align=center]

\pgfplotsset{compat=1.18}
\preprint{CCTP-2026-3 \hfill QMUL-PH-26-08 \\ ITCP-2026-3}

\title{\fontsize{20}{19}\selectfont   
Efficient Conformal Block Evaluation with GoBlocks
}

\author{
    J.~Chryssanthacopoulos\;$^{a,\clubsuit}$, V.~Niarchos\;$^{b,\heartsuit}$, C.~Papageorgakis\;$^{a,\spadesuit}$ and A.~G.~Stapleton\;$^{a,\diamondsuit}$}

\affiliation{
    $^a$ Centre for Theoretical Physics and Astronomy, Department of Physics and Astronomy\\ Queen Mary University of London, London E1 4NS, UK $ $ \\
    $^b$ Department of Physics, ITCP \& CCTP\\
    University of Crete, 71003 Heraklion, Greece\\
    $ $

    {\tt \small
     $^\clubsuit$j.chryssanthacopoulos@qmul.ac.uk, 
     $^\heartsuit$niarchos@physics.uoc.gr,  
     $^\spadesuit$c.papageorgakis@qmul.ac.uk, 
     $^\diamondsuit$a.g.stapleton@qmul.ac.uk}
}

\abstract{\noindent
Conformal blocks in odd spacetime dimensions are not known in closed analytic form. To facilitate efficient computations in the conformal bootstrap, we introduce \texttt{GoBlocks}: a novel
conformal-block generator implemented in the Go programming language, designed for rapid, on-the-fly, parallel evaluation in general spacetime dimensions using recursive relations. The package supports both multi-point and derivative-based bootstrap approaches and allows flexible control over accuracy and performance. We benchmark \gbl against the \sbl package, finding significant speed improvements in applications where computational speed and moderate accuracy are critical, but ultra-high precision  is not essential.
As an illustration, we apply \texttt{GoBlocks} to the mixed-correlator bootstrap of the three-dimensional Ising model, formulated as a non-convex optimisation problem in a suitable truncation scheme.
We simultaneously optimise over external scaling dimensions and OPE CFT data. 
In addition, we discuss how the approach scales as we increase the number of mixed correlators in more general $O(N)$ vector models. 
} 

\date{\today}

\begin{document}

\maketitle

\toc

\section{Introduction and Summary}

The numerical conformal bootstrap has proven to be a powerful non-perturbative tool for rigorously constraining the space of conformal field theories (CFTs) \cite{Rattazzi:2008pe,Poland_2019,Rychkov:2024}. 
When multiple correlators are bootstrapped simultaneously, it can determine CFT data with remarkable accuracy. 
The most celebrated success of this approach is for the four-point bootstrap of the three-dimensional Ising model~\cite{El-Showk:2012cjh,Kos:2014bka,Kos:2016ysd}, which proceeds by scanning over spaces of linear functionals (dual formulation), rather than directly solving the crossing equations (primal formulation). While successful, the linear-functional bootstrap relies crucially on certain positivity assumptions. However, these assumptions are not present in several interesting CFT contexts, e.g., in the presence of defects \cite{Billo:2016cpy,Lauria:2017wav}, boundaries \cite{McAvity:1995zd,Liendo:2012hy}, at finite temperature \cite{Iliesiu:2018fao}, and for the five-point bootstrap \cite{Poland:2023vpn, Poland:2023bny, Poland:2025ide}. This has motivated a revisit of the primal approach through truncations \cite{Gliozzi:2013ysa,Gliozzi:2014jsa,Gliozzi:2015qsa,Gliozzi:2016cmg,Li:2017ukc,Kantor:2021kbx,Kantor:2021jpz, Kantor:2022epi,Li:2023tic,Hu:2025yrs,Huang:2025qkk} and other approximate, deep-learning-based methods \cite{Niarchos:2025cdg}.

At a technical level, both formulations require an efficient computation of conformal blocks. 
In the linear-functional approach, one fixes the dimensions of external operators at each step and employs linear or semidefinite programming tools such as SDPB \cite{Simmons-Duffin:2015qma}. These typically require  expanding the blocks around the crossing-symmetric point---where the OPE converges rapidly---in terms of rational approximations to very high precision and evaluating derivatives up to some order. For the bootstrap of four-point scalar correlators, the de-facto standard is \texttt{scalar\_blocks} \cite{scalar_blocks}, while \texttt{blocks\_3d}  \cite{Erramilli:2020rlr} provides blocks for correlators involving operators in arbitrary  Lorentz representations. These blocks are then fed into semidefinite programming tools, like SDPB, which typically require high numerical precision to avoid numerical instabilities. We note that 
there are also related multi-point approaches where the crossing equations are evaluated on a discrete grid of points in the conformal cross-ratio plane, rather than through derivatives \cite{Hogervorst:2013sma,CastedoEcheverri:2016swn}. 

In primal truncation schemes, one needs to optimise over a high-dimensional space of CFT data that may include the external scaling dimensions of the correlators as dynamical variables. In even dimensions, where conformal blocks are known in closed form, the analytic expressions can be used directly for fast, on-the-fly evaluation. However, in odd dimensions---such as the physically interesting case of 3D---the \texttt{scalar\_blocks} package is prohibitively slow for this purpose and alternative computational tools are needed.\footnote{It should be emphasised that \texttt{scalar\_blocks} is a very efficient tool for the dual formulation of the   numerical conformal bootstrap, where its output can be used directly. For the primal bootstrap, additional processing must be performed on the output, as described below.} Often in these schemes, high numerical accuracy in the conformal-block evaluation is not the primary concern, as there are already other sources  of systematic and stochastic errors that need to be addressed. Despite these shortcomings, there are examples where truncation methods achieve high-accuracy predictions comparable to state-of-the-art results, see e.g.\ \cite{Cavaglia:2021bnz, Cavaglia:2022qpg, Niarchos:2023lot}. 

In this work, we develop efficient methods for conformal-block evaluation 
with generic parameters that include the scaling dimensions of external operators. We present \texttt{GoBlocks}: a fast, parallel, and flexible conformal-block generator implemented in the Go programming language.\footnote{The package can be accessed from GitHub at \href{https://github.com/xand-stapleton/goblocks}{\texttt{https://github.com/xand-stapleton/goblocks}}.} We demonstrate its use in both the multi-point and derivative-based bootstrap of scalar correlation functions in $D$ dimensions. 

As an application, we demonstrate how {\texttt{GoBlocks}} performs in a non-convex optimisation treatment of multiple mixed correlators in the 3D Ising model. Optimisation is performed using the companion package \texttt{BootSTOP-multi-correlator}, a multi-objective implementation of the bootstrap stochastic optimiser, \texttt{BootSTOP} \cite{Kantor:2021jpz, Kantor:2022epi, Kantor:2021kbx, Niarchos:2023lot}. \texttt{GoBlocks} includes a simple Python interface which, in principle, can be used with any optimisation tool independent of \texttt{BootSTOP}. We also analyse the relative performance, stability, and computational trade-offs of the multi-point and derivative strategies.

The rest of this paper is organised as follows. \cref{sec:goblocks} provides an overview of \texttt{GoBlocks}, including its multi-point and derivative formulations. It also discusses benchmarking against \texttt{scalar\_blocks} in terms of accuracy and runtime. \cref{sec:demonstrations} then summarises the results of an implementation of the package to the bootstrap of the
conformal data of the 3D Ising model, and remarks on the scalability to other $O(N)$ vector models. \cref{sec:outlook} concludes with a summary and avenues for future research.
\vskip 0.25cm

\noindent {\it Note added: 
\vskip .25cm
\noindent In v1 of this work, we employed a fixed-point iteration scheme in which the recursion relation~\eqref{eqn:big_daddy_recursive_part_1} is solved by iterating simultaneously over all $r$-orders at each step. This implementation led to observed divergences at large $|\Delta_{ij}|$, $|\Delta_{kl}|$. 
We have now adopted an order-by-order $r$-series expansion, where each Taylor coefficient is computed exactly from previously determined lower-order data. This change eliminates the previous divergences and improves \gbl accuracy  across the parameter space tested, while also leading to speed gains. We would like to thank S.~Rychkov for communication on this point.}

\section{Overview of GoBlocks}
\label{sec:goblocks}

The crossing equations that give rise to the conformal bootstrap programme for four-point correlators depend on the conformal cross-ratio plane, parametrised by $(z,\bar z)$. There are two main approaches for the numerical implementation of the bootstrap in the literature. The most popular method involves computing derivatives of the crossing equations at the crossing-symmetric point, which we call the derivative approach. An alternative method involves working with the crossing equations at specific points in the cross-ratio plane, which we term the multi-point approach \cite{Hogervorst:2013sma,CastedoEcheverri:2016swn}. The numerical optimisation of CFT data in both approaches relies on the accurate evaluation of conformal blocks. 

The state-of-the-art method for the calculation of scalar conformal blocks is \texttt{scalar\_blocks}, an efficient C++ implementation of the Zamolodchikov recursion relations capable of producing conformal blocks with exceptional precision \cite{scalar_blocks}. This package computes derivatives of conformal blocks at the crossing-symmetric point $z=\bar{z}=1/2$:
\begin{equation}
    \label{eq:scalar_blocks_output}
    \partial_z^m\partial^n_{\bar{z}}g^{\Delta_{ij},\Delta_{kl}}_{\Delta,\ell}(z,\bar{z})\vert_{(z,\bar{z})=\frac{1}{2}} \approx \frac{r_0^\Delta}{Q_{\kappa,\ell}(\Delta)}P_{\ell,\kappa,N;m,n}^{\Delta_{ij},\Delta_{kl}}(x),
\end{equation}
where $\Delta_{ij}\coloneqq \Delta_i-\Delta_j$ and $\Delta_{kl}\coloneqq \Delta_k-\Delta_l$ denote differences in scaling dimensions of scalar operators in the four-point function $\langle\mathcal{O}_i\mathcal{O}_j\mathcal{O}_k\mathcal{O}_l\rangle$. The subscripts $N$ and $\kappa$ denote the order of the $r$ expansion and pole order, whilst $\ell$ and $\Delta$ represent the spin and dimension of the exchanged operator. The functions $Q(\Delta)$ and $P(x)$ are polynomials, with $x\coloneqq \Delta - (\ell + D - 2)$, and $r_0 \coloneqq 3 - 2\sqrt{2}$ the crossing-symmetric point in radial coordinates. This rational form allows evaluation at arbitrary $\Delta$ without rerunning the recursion. 

Three important parameters in \texttt{scalar\_blocks} are \texttt{max-derivs}, \texttt{order}, and \texttt{poles}. The parameter \texttt{max-derivs} encodes $n_\text{max}$, defined such that $m+n\leq 2n_\text{max}-1$. The parameter \texttt{order}, given by $N$ above,
sets the maximum order of the $r$ expansion along the $z=\bar{z}$ diagonal. Similarly, the parameter \texttt{poles} specifies how many poles to retain among those whose residues start with $r^n$, $n \leq \kappa$. There is also a \texttt{precision} parameter, which controls the numerical precision of the output and calculations internal to \texttt{scalar\_blocks}. In this work, \texttt{precision} is set to 64 unless otherwise stated. The package returns the expansion coefficients of 
$P$ and the poles $\Delta_i$, which can be used to construct $P(x)$ and $Q(\Delta)$, rather than the derivatives of $g$ directly.

In many applications, one needs to vary dynamically over the unknown scaling dimensions of the external operators, entering as extra parameters in the blocks. This requires direct, repeated application of \texttt{scalar\_blocks}, which is typically prohibitively slow. Since \texttt{scalar\_blocks} returns the ingredients to assemble the conformal blocks, additional processing needs to be performed to recover the blocks themselves, introducing computational overhead.\footnote{The package \texttt{BootSTOP-multi-correlator} includes an efficient \texttt{JAX} implementation of the necessary processing steps.}

To circumvent this difficulty, a potential alternative approximation of the blocks could use multilinear or polynomial interpolation. These approaches pre-compute block values on a fixed grid of scaling dimensions and interpolate to estimate values at arbitrary points. Whilst interpolation is computationally efficient, its accuracy for highly non-linear functions, such as the conformal blocks, is limited by the grid resolution, and increasing the resolution to improve accuracy substantially increases memory requirements. For a quantitative assessment of multilinear and polynomial interpolation accuracy, see \cref{app:multilinear_polynomial_interpolation}.

Fortunately, recursive relations for evaluating conformal blocks can be iterated efficiently provided they are implemented in a sufficiently performant language. To that end, we developed \texttt{GoBlocks}, an efficient, parallel, multi-threaded block generator written in Go. The current implementation of \texttt{GoBlocks} supports both multi-point and derivative approaches, outlined in \cref{sec:pts_approach,sec:deriv_approach}, respectively.

\subsection{Multi-point Approach}
\label{sec:pts_approach}

For the implementation of the multi-point approach, we work in radial coordinates $(r, \eta)$, which are related to the cross-ratios by
\begin{equation}
    re^{i\theta} = \frac{z}{(1+\sqrt{1-z})^2},\quad \eta = \cos\theta.
\end{equation}
The conformal blocks in these coordinates can be written as
\begin{equation}
  \label{eqn:big_daddy_recursive_part_1}
  h_{\Delta, \ell}^{\Delta_{ij}, \Delta_{kl}}  (r, \eta) = r^{-\Delta}g_{\Delta, \ell}^{\Delta_{ij}, \Delta_{kl}}  (r, \eta)
  =
  \tilde{h}_{\ell}^{\Delta_{ij}, \Delta_{kl}}  (r, \eta) + \sum_{i}
  \frac{c_i^{\Delta_{ij}, \Delta_{kl}}}{\Delta-\Delta_i} r^{n_i}
  h_{\Delta_i + n_i, \ell_i}^{\Delta_{ij}, \Delta_{kl}}  (r, \eta),
\end{equation}
where $\Delta_i$ are the positions of the poles with associated data $n_i$, $\ell_i$, and $c_i^{\Delta_{ij}, \Delta_{kl}}$. There are three types of poles, with the corresponding data summarised in Table 1 of \cite{Kos:2014bka}. The function $\tilde{h}$ seeds the recursion with initial data, and can be written in terms of a Gegenbauer polynomial. See Eq.~(4.6) of \cite{Kos:2014bka} for a complete expression. To solve \cref{eqn:big_daddy_recursive_part_1}, we expand $h$ in a Taylor series in $r$ around $r=0$ and compute the coefficients order by order. At each order $p$, the contribution from a pole with shift $n_i$ uses only coefficients at order $p - n_i < p$, which have already been determined exactly. This ensures that no approximate higher-order terms are generated. The blocks are returned in terms of the following linear combinations, which appear directly in the crossing equations:
\begin{equation}
    \label{eq:F_definition}
    F_{\pm,\Delta,\ell}^{ij,kl}(u, v) \coloneqq v^{\frac{\Delta_j + \Delta_k}{2}}g_{\Delta,\ell}^{\Delta_{ij},\Delta_{kl}}(u,v) \pm u^{\frac{\Delta_j + \Delta_k}{2}}g_{\Delta,\ell}^{\Delta_{ij},\Delta_{kl}}(v,u),
\end{equation}
where $u\coloneqq z\bar{z}$ and $v\coloneqq (1-z)(1-\bar{z})$. These functions depend on four scaling dimensions: $\Delta_{ij}$, $\Delta_{kl}$, $\bar{\Delta}_{jk}\coloneqq (\Delta_j + \Delta_k)/2$, and $\Delta$.

The multi-point algorithm consists of three steps: recursion, block evaluation, and construction of $F_\pm$. These steps are depicted in \cref{fig:multipoint_flowchart} of \cref{app:schematic_overview}. In the recursion step, the blocks $h_{\Delta,\ell}^{\Delta_{ij},\Delta_{kl}}$ are computed at a set of pole locations given the external operator dimensions $\Delta_{ij}$ and $\Delta_{kl}$. This is the most computationally intensive step. The evaluation step then accepts a list of spins and scaling dimensions of exchanged operators, and uses the recurse-step results to compute the value of $g$ for each operator. In the construction step, the $F_\pm$ are constructed from \cref{eq:F_definition}. Since \cref{eq:F_definition} requires both $g(u,v)$ and $g(v,u)$, the recursion step needs to be run twice with different arguments. The algorithm returns $F_\pm$ at the set of points in the $z$ plane that were requested.

 \cref{tab:goblocks_parameters} summarises the parameters used by \texttt{GoBlocks} in the multi-point approach. The parameters \texttt{k\_1\_max} and \texttt{k\_2\_max} set the maximum number of Type I and Type II poles for recursion, analogous to the \texttt{poles} parameter in \texttt{scalar\_blocks}. The \texttt{max\_iterations} parameter sets the maximum Taylor order in the $r$-series expansion,\footnote{More specifically, the order is determined by $\text{max}(2\,\times$~\texttt{k\_2\_max}, \texttt{max\_iterations}$)$, which ensures all pole types have contributed.} 
 playing a role analogous to the \texttt{order} parameter in \texttt{scalar\_blocks}. The \texttt{tol} parameter enables early termination: if the maximum fractional contribution of the latest Taylor term falls below \texttt{tol}, the series is considered converged. 
 To ensure robust convergence, the early termination check is deferred until a sufficient number of terms have been computed.
 The parameter \texttt{ell\_max} sets the maximum spin included in the recursion; for optimal accuracy, it should be set larger than the highest spin of interest, since poles of spin $\ell$ reference spins up to $\ell + k_{1}^{\text{max}}$.

\begin{table}[bt]
\centering
\small
\caption{Configuration parameters of \texttt{GoBlocks} using the multi-point and derivative approaches. The derivative approach shares all the same parameters as the points approach, but has additional parameters controlling the number of derivatives computed and returned.}

\label{tab:goblocks_parameters}
\begin{tabular}{lll}
\toprule
\textbf{Parameter} & \textbf{Approach} & \textbf{Description} \\
\midrule
\texttt{k\_1\_max} & Points, derivatives & Maximum order of Type I poles \\
\texttt{k\_2\_max} & Points, derivatives & Maximum order of Type II poles \\
\texttt{ell\_min}  & Points, derivatives & Minimum spin in the recursion \\
\texttt{ell\_max}  & Points, derivatives & Maximum spin in the recursion \\
\texttt{max\_iterations} & Points, derivatives & Maximum Taylor order ($r$-series) \\
\texttt{tol} & Points & Tolerance for early termination \\
\texttt{n\_max} & Derivatives & Maximum derivative order \\
\texttt{num\_derivs\_to\_keep} & Derivatives & Number of derivatives to return \\
\bottomrule
\end{tabular}
\end{table}

\subsection{Derivative Approach}
\label{sec:deriv_approach}
An alternative to the multi-point approach is to use derivatives of the conformal blocks at the crossing-symmetric point, as in \texttt{scalar\_blocks}.   
\cref{app:recursive_deriv} derives a recursive formula for block derivatives with respect to $(r,\eta)$ at the crossing-symmetric point. Converting to derivatives with respect to $(z,\bar{z})$ requires derivatives of $r$ and $\eta$ with respect to $(z,\bar{z})$, which are derived in \cref{app:z_zb_r_eta_rosetta}. The derivative approach in \gbl returns derivatives of $F_\pm$. From \cref{eq:F_definition}, these derivatives can be written in terms of derivatives of $g$ as
\begin{equation}
    \label{eq:F_derivatives}
    \partial_z^m \partial_{\bar{z}}^n F_{\pm,\Delta,\ell}^{ij,kl}(z, \bar{z}) = \left(1 \pm (-1)^{m+n}\right)\sum_{p=0}^m \sum_{q=0}^n \binom{m}{p}
        \binom{n}{q}c_{pq}^{\bar{\Delta}_{jk}}\partial_z^{m-p}\partial_{\bar{z}}^{n-q} g_{\Delta,\ell}^{\Delta_{ij},\Delta_{kl}}(z, \bar{z}),
\end{equation}
where $c_{ij}^\alpha \coloneqq (-1)^{i+j}2^{i+j-2\alpha}(1+\alpha-i)_i (1+\alpha-j)_j$, with $(x)_n$ the rising factorial. For the same number of constraints (i.e. points and derivatives), the runtimes of the multi-point and derivative approaches are comparable when run with identical shared hyperparameters (e.g. $\ell_\text{max}, k_{1,2}^\text{max}$).\footnote{Empirically, for the same number of constraints, running on x64 architectures appears to give a slight (approximately 25\%) performance advantage to the multi-point method over the derivative approach. This has not been observed on ARM architectures running MacOS.}

The algorithmic flow of the derivative approach is depicted in \cref{fig:deriv_flowchart} of \cref{app:schematic_overview}. The recursion and evaluation steps are very similar to the multi-point approach. In the construction of the $F_\pm$ derivatives, there is an additional step to convert from derivatives of $g$ with respect to $(r,\eta)$ to derivatives with respect to $(z, \bar{z})$. These steps involve combinatorial sums whose cost grows steeply with \texttt{n\_max} and introduce the dominant computational cost in the derivative approach. To mitigate this, the required combinatorial factors are precomputed and cached, so that subsequent evaluations at different operator dimensions are fast. \texttt{GoBlocks} also supports GPU acceleration via CUDA for the matrix operations in the conversion step, which typically becomes advantageous when \texttt{n\_max} is large.

\cref{tab:goblocks_parameters} summarises the parameters in the derivative approach. This method shares all parameters with the multi-point approach but introduces two new controls: the maximum derivative order, \texttt{n\_max}, and the number of derivatives retained, \texttt{num\_derivs\_to\_keep}. The maximum derivative order, \texttt{n\_max}, is consistent with its usage in \texttt{scalar\_blocks}. The parameter \texttt{num\_derivs\_to\_keep} governs the subset of computed derivatives that are returned, directly impacting numerical accuracy. By selecting a smaller subset for a fixed \texttt{n\_max}, the retained derivatives exhibit higher precision compared to retaining a larger set.

\subsection{Accuracy}
\label{sec:accuracy}

We next present a comparison of the accuracy of \texttt{GoBlocks} and \texttt{scalar\_blocks} over a range of scaling dimensions. The output from \texttt{scalar\_blocks}, computed with large parameter values, serves as the reference for evaluating the accuracy of \texttt{GoBlocks}. The parameters used for \texttt{scalar\_blocks} were \texttt{order} = 90, \texttt{precision} = 90, \texttt{poles} = 20, and \texttt{max-derivs} = 14. In practice, it is preferable to compare the output of \texttt{GoBlocks} and \texttt{scalar\_blocks} using $F_{\pm,\Delta,\ell}$ rather than $g_{\Delta,\ell}$ directly. This approach reduces noise in the estimates, especially when $g$ is small, and provides smoother results for $F$ and its derivatives.

\paragraph{Multi-point tests:}\cref{tab:z-points-errors} presents the percent errors of the \texttt{GoBlocks} multi-point approach relative to \texttt{scalar\_blocks} for various parameters. The comparison is performed at the crossing-symmetric point $z=\bar{z}=1/2$, the only non-derivative term returned by \texttt{scalar\_blocks}, corresponding to  $m=n=0$. The results focus on $F_{+,\Delta,\ell}$ from both methods. The relative error is defined as the difference between the two values, normalised by the value of $F_{+,\Delta,\ell}$ from \texttt{scalar\_blocks}. For each spin up to six, $1{,}000$ points in the $(\Delta_{ij},\Delta_{kl},\bar{\Delta}_{jk}, \Delta)$ space were evaluated with $|\Delta_{ij}|,|\Delta_{kl}| \leq 3.5$. The table reports the error percentiles and the mean. Across all parameters tested, the errors remain well below a tenth of a percent. Increasing $\ell_\text{max}$ significantly improves performance, but accuracy is relatively insensitive to $k_{1,2}^\text{max}$.

\begin{table}[bt]
\centering
\small
\begin{tabular}{cc ccccc}
\toprule
\multicolumn{2}{c}{\textbf{Parameters}} &
\multicolumn{5}{c}{\textbf{Error (\%)}} \\
\cmidrule(lr){1-2} \cmidrule(lr){3-7}
$k_{1,2}^\text{max}$ & $\ell_{\text{max}}$ &
25\% & 50\% & 75\% & 90\% & Mean \\
\midrule
10 & 10 & $5.517\times 10^{-6}$  & $1.875\times 10^{-4}$ & $9.039\times 10^{-3}$ & $1.241\times 10^{-1}$ & $8.157\times 10^{-2}$ \\
10 & 16 & $1.220\times 10^{-8}$  & $4.839\times 10^{-8}$ & $1.159\times 10^{-6}$ & $1.837\times 10^{-5}$ & $2.077\times 10^{-5}$ \\
10 & 20 & $4.588\times 10^{-9}$  & $1.165\times 10^{-8}$ & $2.744\times 10^{-8}$ & $6.375\times 10^{-8}$ & $4.190\times 10^{-8}$ \\
16 & 10 & $5.517\times 10^{-6}$  & $1.875\times 10^{-4}$ & $9.039\times 10^{-3}$ & $1.241\times 10^{-1}$ & $8.157\times 10^{-2}$ \\
16 & 20 & $1.687\times 10^{-12}$  & $1.520\times 10^{-11}$ & $1.174\times 10^{-9}$ & $1.914\times 10^{-8}$ & $2.539\times 10^{-8}$ \\
20 & 20 & $1.551\times 10^{-12}$  & $1.505\times 10^{-11}$ & $1.174\times 10^{-9}$ & $1.914\times 10^{-8}$ & $2.539\times 10^{-8}$ \\
\bottomrule
\end{tabular}

\caption{Percent errors of the \texttt{GoBlocks} multi-point approach relative to \texttt{scalar\_blocks} for different parameters, defined in \cref{tab:goblocks_parameters}. Statistics are aggregated across all spins up to six and for $1{,}000$ points in the $(\Delta_{ij},\Delta_{kl},\bar{\Delta}_{jk}, \Delta)$ plane with $|\Delta_{ij}|,|\Delta_{kl}|\leq 3.5$.}
\label{tab:z-points-errors}
\end{table}
 
As a complementary validation, we also compared the output of \texttt{GoBlocks} against the closed-form ${}_3F_2$ expression for conformal blocks on the diagonal $z = \bar{z}$, which is valid  for $\Delta_{ij} = 0$ and was derived in \cite{Hogervorst:2013sma}. This provides an independent cross-check using an exact analytic result.\footnote{The two methods use different normalisation conventions, related by $F_{\pm}^{\texttt{GoBlocks}} = (-1)^\ell \, r_0^{\,\Delta} \, F_{\pm}^{{}_{3}F_{2}}$. All errors were computed after applying this rescaling.}  We evaluated $F_{\pm,\Delta,\ell}$ for a grid of 36 parameter combinations: three values of $\bar{\Delta}_{jk}$ while keeping $\Delta_{ij} = 0$ (including the equal external-dimension case $\Delta_i = \Delta_j = \Delta_k = \Delta_l$), four spins $\ell \in  \{0, 1, 2, 4\}$, and three internal operator dimensions $\Delta \in \{1.5, 2.5, 4.5\}$, each at nine diagonal points in the range $z \in [0.3, 0.7]$. All 324 comparisons for $F_+$ and all 288 non-vanishing comparisons for $F_-$ achieve relative errors below $10^{-5}$, with median errors of $5.0 \times 10^{-9}$ and $2.4 \times 10^{-8}$, respectively.
The error scales with the distance from the crossing-symmetric point, ranging from $\sim\!10^{-10}$ at $z = 1/2$ to $\sim\!10^{-7}$ at $|z - 1/2| = 0.2$, and increases modestly with spin, from $\sim\!10^{-10}$ at $\ell = 0$ to $\sim\!10^{-8}$ at $\ell = 4$.

As a further validation beyond the diagonal, we compared the multi-point output of \texttt{GoBlocks} against a high-precision Mathematica implementation of the same Zamolodchikov recursion, which uses 200-digit working precision and the order-by-order $r$-series approach. This comparison covers complex $z$-points (with $\bar{z} = z^*$) and non-zero $\Delta_{ij}<3$, $\Delta_{kl}<3$, regimes inaccessible to the ${}_3F_2$ formula. Across a number of test points spanning the OPE convergence region and a range of  $\Delta_{ij}$, $\Delta_{kl}$, all comparisons with $\ell_{\text{max}} = 10$ achieve relative errors below $10^{-3}$. Extending to $\ell_{\text{max}} = 20$ brings the most demanding tested cases (of simultaneous $\Delta_{ij}=3$ and $\Delta_{kl} = 3$ at $z = 0.7$) to $\sim\!4 \times 10^{-7}$.

\paragraph{Derivative tests: } Table \ref{tab:derivative-errors} reports percent errors of the \texttt{GoBlocks} derivative approach relative to \texttt{scalar\_blocks} for different parameters. In this case, the errors are aggregated across $\partial_z\partial_{\bar{z}}F_{+,\Delta,\ell}$ and $\partial_z\partial_{\bar{z}}F_{-,\Delta,\ell}$. At $n_\text{max}=8$, $n_\text{deriv}=21$ derivatives can be supported at high accuracy. The error beyond the 21st derivative increases considerably. At $n_\text{max}=9$, $28$ derivatives can be reasonably supported.
Although the performance of the derivative approach is very good overall, it is inferior to the multi-point approach. Increasing $n_\text{max}$ while keeping the other parameters fixed allows more derivatives to be captured, but generally comes at the price of worse performance overall.

\begin{table}[bt]
\centering
\footnotesize
\begin{tabular}{cccc ccccc}
\toprule
\multicolumn{4}{c}{\textbf{Parameters}} &
\multicolumn{5}{c}{\textbf{Error (\%)}} \\
\cmidrule(lr){1-4} \cmidrule(lr){5-9}
$k_{1,2}^\text{max}$ & $\ell_{\text{max}}$ & $n_{\text{max}}$ & $n_{\text{deriv}}$ &
25\% & 50\% & 75\% & 90\% & Mean \\
\midrule
10 & 16  & 8 & 21 & $4.748\times10^{-5}$ & $5.000\times10^{-4}$ & $4.000\times10^{-3}$ & $2.570\times10^{-2}$ & $9.730\times10^{-2}$ \\
10 & 16  & 9 & 28 & $1.034\times10^{-4}$ & $1.200\times10^{-3}$ & $9.300\times10^{-3}$ & $5.710\times10^{-2}$ & $2.142\times10^{-1}$ \\
10 & 20  & 8 & 21 & $8.518\times10^{-6}$ & $7.437\times10^{-5}$ & $5.963\times10^{-4}$ & $3.731\times10^{-3}$ & $1.989\times10^{-2}$ \\
10 & 20  & 9 & 28 & $1.837\times10^{-5}$ & $1.839\times10^{-4}$ & $1.571\times10^{-3}$ & $1.008\times10^{-2}$ & $6.874\times10^{-2}$ \\
16 & 20  & 8 & 21 & $1.463\times10^{-7}$ & $3.310\times10^{-6}$ & $5.206\times10^{-5}$ & $5.614\times10^{-4}$ & $3.754\times10^{-3}$ \\
16 & 20  & 9 & 28 & $4.244\times10^{-7}$ & $9.964\times10^{-6}$ & $1.553\times10^{-4}$ & $1.574\times10^{-3}$ & $2.365\times10^{-2}$ \\
20 & 20  & 8 & 21 & $8.202\times10^{-8}$ & $2.657\times10^{-6}$ & $4.804\times10^{-5}$ & $5.000\times10^{-4}$ & $3.800\times10^{-3}$ \\
20 & 20  & 9 & 28 & $2.547\times10^{-7}$ & $8.003\times10^{-6}$ & $1.403\times10^{-4}$ & $1.500\times10^{-3}$ & $2.370\times10^{-2}$ \\
25 & 25  & 8 & 21 & $1.827\times10^{-10}$ & $6.949\times10^{-9}$ & $1.869\times10^{-7}$ & $2.787\times10^{-6}$ & $2.1283\times10^{-5}$ \\
25 & 25  & 9 & 28 & $6.346\times10^{-10}$ & $2.628\times10^{-8}$ & $7.267\times10^{-7}$ & $1.152\times10^{-5}$ & $1.913\times10^{-4}$ \\
\bottomrule
\end{tabular}
\caption{Percent errors of the \texttt{GoBlocks} derivative approach relative to \texttt{scalar\_blocks} for different parameters, defined in \cref{tab:goblocks_parameters}. The statistics are aggregated across all spins up to six and for $1{,}000$ points in the $(\Delta_{ij},\Delta_{kl},\bar{\Delta}_{jk}, \Delta)$ plane with $|\Delta_{ij}|,|\Delta_{kl}|\leq3.5$. The statistics are also aggregated over errors for both $F_+$ and $F_-$.}
\label{tab:derivative-errors}
\end{table}

\subsection{Timing}
\label{sec:timing}

In numerical computations, there is a natural trade-off between speed and accuracy. The higher the accuracy one wishes to achieve, the longer the block-evaluation time. Quantifying this trade-off involves measuring accuracy as a function of runtime, which requires a robust reference standard. In this work, the benchmark is defined by the output of \sbl evaluated at an \texttt{order} of $200$. This high-precision result is treated as ground truth for systematic comparison, enabling both \sbl and the \gbl derivative approach to be evaluated across a range of parameter settings. All computations were performed on an Intel Core i5-12400 processor running Arch Linux with kernel version 6.17.

Each evaluation involved sampling $100$ points in the $(\Delta_{ij},\Delta_{kl},\bar{\Delta}_{jk}, \Delta)$ parameter space with six different spins. Parameters shared by both algorithms were set to the same value. For example, \texttt{poles} in \texttt{scalar\_blocks} was matched to \texttt{k\_1\_max} and \texttt{k\_2\_max} in \texttt{GoBlocks}, and \texttt{max-derivs} was matched to \texttt{n\_max} and \texttt{num\_derivs\_to\_keep}. The main parameters varied were \texttt{order} in \texttt{scalar\_blocks}, and \texttt{ell\_max} and \texttt{tol} in \texttt{GoBlocks}. 

\cref{fig:scalar_blocks_runtime_vs_order} presents the runtime of \sbl as a function of the \texttt{order} parameter, with the shaded region indicating the standard deviation across multiple runs. As described previously, \sbl outputs expansion coefficients for polynomials centred at the crossing-symmetric point. To obtain the final block derivatives, further processing is required. Derivatives of $g$ are constructed using \cref{eq:scalar_blocks_output}, and then converted to derivatives of $F_\pm$ via \cref{eq:F_derivatives}. This post-processing was implemented in Python using a multicore \texttt{JAX} backend. The figure displays runtimes both with and without this post-processing step. Without post-processing, \sbl produces expansion coefficients in under a second for \texttt{order} up to $90$. Including post-processing, runtimes increase substantially, making real-time block evaluation impractical for some applications. In subsequent runtime reports for \texttt{scalar\_blocks}, only the unprocessed output times are shown, representing the absolute lower bound for evaluation time.

\begin{figure}[bt]
    \centering
    \begin{tikzpicture}
\begin{axis}[
    xlabel={Order},
    ylabel={Runtime (s)},
    xmode=linear,
    ymode=linear,
    legend pos=north west,
    legend style={
        at={(0.2, 0.95)},
        anchor=north,
    },
    ymajorgrids=true,
    xmajorgrids=true,
]

\addplot[
    thick,
    blue,
    mark=*,
    mark options={solid}
] coordinates {
    (10,0.05308466863632203)
    (12,0.06152029848098756)
    (14,0.06364499092102051)
    (16,0.06598081493377686)
    (18,0.06881218910217286)
    (20,0.07196539878845215)
    (30,0.1005992360115051)
    (40,0.1675230550765991)
    (50,0.2730476279258728)
    (60,0.4229439015388489)
    (70,0.6386459317207336)
    (80,0.9419349064826967)
    (90,1.350334534645081)
};
\addlegendentry{Raw}

\addplot[
    thick,
    red,
    mark=*,
    mark options={solid}
] coordinates {
    (10,0.8164500250816346)
    (12,0.8247375841140746)
    (14,0.8229440956115721)
    (16,0.8218301372528077)
    (18,0.8267693581581116)
    (20,0.8288212771415711)
    (30,0.8562935576438905)
    (40,0.9229045233726501)
    (50,1.028644933223724)
    (60,1.178508754253388)
    (70,1.39513594865799)
    (80,1.698948146820068)
    (90,2.106255970478058)
};
\addlegendentry{Total}

\addplot[
    fill=red,
    fill opacity=0.2,
    draw=none
] coordinates {
    (10,0.8311985239157206)
    (10,0.8017015262475485)
    (12,0.8072635688521406)
    (14,0.8052334734158568)
    (16,0.8034882500128611)
    (18,0.807696564386533)
    (20,0.81000146253398)
    (30,0.8285571092351671)
    (40,0.8842573963306989)
    (50,0.9841839979305813)
    (60,1.12254827398588)
    (70,1.323355386174053)
    (80,1.608157608320532)
    (90,1.990961373501027)
    (90,2.221550567455089)
    (90,2.221550567455089)
    (80,1.789738685319604)
    (70,1.466916511141927)
    (60,1.234469234520895)
    (50,1.073105868516867)
    (40,0.9615516504146012)
    (30,0.8840300060526138)
    (20,0.8476410917491621)
    (18,0.8458421519296901)
    (16,0.8401720244927544)
    (14,0.8406547178072874)
    (12,0.8422115993760086)
    (10,0.8311985239157206)
    (10,0.8311985239157206)
};


\addplot[
    fill=blue,
    fill opacity=0.2,
    draw=none
] coordinates {
    (10,0.05836463448110647)
    (10,0.0478047027915376)
    (12,0.05327392097093282)
    (14,0.054740568767189)
    (16,0.05659168294069571)
    (18,0.05856810017807798)
    (20,0.061369938098157)
    (30,0.08117666704718561)
    (40,0.1372772736481453)
    (50,0.2373694768977334)
    (60,0.375845847623156)
    (70,0.5757448380950629)
    (80,0.8602994107934947)
    (90,1.244377516039603)
    (90,1.456291553250558)
    (90,1.456291553250558)
    (80,1.023570402171899)
    (70,0.7015470253464042)
    (60,0.4700419554545418)
    (50,0.3087257789540122)
    (40,0.197768836505053)
    (30,0.1200218049758246)
    (20,0.08256085947874731)
    (18,0.07905627802626775)
    (16,0.07536994692685801)
    (14,0.07254941307485201)
    (12,0.06976667599104229)
    (10,0.05836463448110647)
    (10,0.05836463448110647)
};

\end{axis}
\end{tikzpicture}
    \caption{Runtime of \sbl as \texttt{order} varies, with and without post-processing to construct block derivatives and convert to derivatives of $F_\pm$. Without post-processing, \sbl runs in under a second for most \texttt{order} values. With post-processing, the runtime can increase to two seconds.}
    \label{fig:scalar_blocks_runtime_vs_order}
\end{figure}

\cref{fig:goblocks_accuracy_vs_parameters} presents the performance of \gbl for varying hyperparameter configurations for $n_\text{max}=8$ for block derivatives.
The parameters $k_{1,2}^\text{max}$ and $\ell_\text{max}$ have the largest impact on overall accuracy. 
\cref{fig:goblocks_dev_vs_ellmax_k12-20} shows the average associated deviation from ground truth for varied $\ell_\text{max}$ and fixed $k_{1,2}^\text{max}=20$. 
The deviation is computed as the $L^2$ norm of the element-wise difference between the \gbl results and the benchmark, normalised by the benchmark.
Similarly, \cref{fig:goblocks_runtime_vs_ellmax_k12-20} presents the average runtime of \gbl for varied $\ell_\text{max}$ and fixed $k_{1,2}^\text{max}=20$. We stress that these configurations do not yield the best accuracy-to-runtime proposition, but are instead intended to present an example of how $\ell_\text{max}$ affects runtime and block accuracy.

\begin{figure}[bt]
    \centering
    \begin{subfigure}{0.45\textwidth}
        \centering




\begin{tikzpicture}
\begin{axis}[
    xlabel={$\ell_\text{max}$},
    ylabel={Error (\%)},
    ymode=log,
    xmode=linear,
    xmin=10,
    xmax=40,
    ymin=0.000001,
    enlarge x limits=false,
    ymajorgrids=true,
    xmajorgrids=true,
    width=\textwidth,
    height=0.8\textwidth,
    ymin=1e-5,ymax=1e3,
    ytick={1e-5,1e-3,1e-1,1e1,1e3},
    yticklabel style={/pgf/number format/fixed}
]

\addplot[
    color=blue,
    thick,
    mark=*,
] table {
x y
10 21.44784332573506
12 3.920892455981622
14 0.8556203873152105
16 0.1938545096882345
18 0.03573741918211738
20 0.00550639109425956
25 4.425198783481717e-05
30 6.603715693863736e-05
35 6.595951572608099e-05
40 6.595965966050203e-05
};


\end{axis}
\end{tikzpicture}
        \vspace{-1.5em}
        \caption{}
        \label{fig:goblocks_dev_vs_ellmax_k12-20}
    \end{subfigure}
    \hspace{1.5em}
    \begin{subfigure}{0.45\textwidth}
        \centering
        \begin{tikzpicture}
\begin{axis}[
    xlabel={$\ell_\text{max}$},
    ylabel={Runtime (s)},
    ymode=linear,
    xmode=linear,
    ymajorgrids=true,
    xmajorgrids=true,
    width=\textwidth,
    height=0.8\textwidth,
    ymin=0,ymax=0.3,
    ytick={0,0.1,0.2,0.3},
    xmin=10,xmax=35,
    xtick={10,15,20,25,30,35},
    yticklabel style={/pgf/number format/fixed}
]

\addplot[
    color=blue,
    thick,
    mark=*,
] table {
x y
12 0.06894121551513671
13 0.0760519428253174
14 0.08358182668685914
15 0.09121609354019164
16 0.1003673934936523
17 0.1096421885490418
18 0.1197665729522705
19 0.1307935094833374
20 0.1421044111251831
25 0.2019111280441284
30 0.2675233459472657
};

\addplot[
    fill=blue,
    fill opacity=0.2,
    draw=none
] coordinates {
(12,0.06942408043021224)
(12,0.06845835060006118)
(13,0.07560881904334522)
(14,0.08306118305880764)
(15,0.09058711650614173)
(16,0.09949448372077034)
(17,0.1088539361577232)
(18,0.1189977300920977)
(19,0.1299276086011923)
(20,0.1411409387828275)
(25,0.2006018365795451)
(30,0.2642734001934852)
(30,0.2707732917010461)
(30,0.2707732917010461)
(25,0.2032204195087117)
(20,0.1430678834675387)
(19,0.1316594103654825)
(18,0.1205354158124433)
(17,0.1104304409403603)
(16,0.1012403032665343)
(15,0.09184507057424154)
(14,0.08410247031491064)
(13,0.07649506660728957)
(12,0.06942408043021224)
(12,0.06942408043021224)
};

\end{axis}
\end{tikzpicture}
        \vspace{-1.5em}
        \caption{}
        \label{fig:goblocks_runtime_vs_ellmax_k12-20}
    \end{subfigure}
    \caption{
    Accuracy and runtime of \gbl with increasing $\ell_\text{max}$.
    Fig.~(a) shows block accuracy varying from $\sim 20\%$ to $\sim 4 \times 10^{-5}\%$ as $\ell_\text{max}$ increases from $12$ to $30$ for fixed $k_{1,2}^\text{max}=20$. Fig.~(b) shows runtime as a function of $\ell_\text{max}$ for fixed $k_{1,2}^\text{max}=20$.}
    \label{fig:goblocks_accuracy_vs_parameters}
\end{figure}

\cref{fig:average_runtime_vs_accuracy} compares the average runtime of \sbl and \gbl at fixed block accuracy. In practice, to achieve the optimal accuracy-to-speed ratio, the parameters $k_{1,2}^\text{max}$ and $\ell_\text{max}$ must be tuned. To achieve the results stated in \cref{fig:average_runtime_vs_accuracy}, parameter values $k_{1,2}^\text{max}=10$ and $\ell_\text{max}=16$ were used. Up to the $10^{-4}\%$ accuracy level, we were able to find combinations of the triple $k_{1,2}^\text{max}, \ell_\text{max}$ which were approximately $4.6$ times faster than \texttt{scalar\_blocks}. Hyperparameter searches for higher accuracy levels were not performed. \cref{app:tuning_ktriple} presents additional information on parameter tuning.
The reported \gbl runtime includes the additional step of converting to derivatives of $F_\pm$, while the \sbl runtime reflects only the generation of raw output. Applications requiring fully constructed derivatives of $g$ or $F_\pm$ would incur further significant processing time beyond what is shown for \texttt{scalar\_blocks}. 

\begin{figure}
    \centering
    \begin{tikzpicture}
\begin{axis}[
    ybar,
    bar width=8pt,
    width=11cm,
    height=6.5cm,
    ylabel={Runtime (seconds)},
    xlabel={Target Deviation Level (\%)},
    xlabel style={yshift=-0.6em},
    symbolic x coords={0.1\%,0.5\%,1\%,5\%,10\%,20\%},
    xtick=data,
    ymin=0,
    ymax=0.22,
    legend cell align={left},
    legend pos=north west,
    legend style={
        at={(1.25,0.5)},
        anchor=north,
        column sep=3pt,
        draw=none,
    }
]

\addplot+[
    error bars/.cd,
    y dir=both,
    y explicit
] coordinates {
(0.1\%,0.1635109696336972) +- (0,0.0338729235909236)
(0.5\%,0.1618294822926424) +- (0,0.0338077575828667)
(1\%,0.1618294822926424) +- (0,0.0338077575828667)
(5\%,0.1618294822926424) +- (0,0.0338077575828667)
(10\%,0.1611905183792114) +- (0,0.0337674264073696)
(20\%,0.1611905183792114) +- (0,0.0337674264073696)
};

\addplot+[
    error bars/.cd,
    y dir=both,
    y explicit
] coordinates {
(0.1\%,0.03561570921609569) +- (0,0.00071996546869609)
(0.5\%,0.03552642729547289) +- (0,0.00072779288978312)
(1\%,0.03549475615051971) +- (0,0.00070464133479835)
(5\%,0.03547984983610070) +- (0,0.00072344152450638)
(10\%,0.03545809934536616) +- (0,0.00073979955058162)
(20\%,0.03545383374715588) +- (0,0.00073716200641520)
};

\legend{\texttt{scalar\_blocks}, \texttt{goblocks}}

\end{axis}
\end{tikzpicture}
    \caption{Comparison of average runtimes between \sbl and \gbl for given accuracies with $k_{1,2}^\text{max}=10$. Overall, \gbl is approximately $4.6$ times faster than \sbl between the 0.1\% and 10\% accuracy levels.}
    \label{fig:average_runtime_vs_accuracy}
\end{figure}

\section{Applications}
\label{sec:demonstrations}

We will next demonstrate the utility of \texttt{GoBlocks} in conformal bootstrap studies with multiple correlators, starting with the 3D Ising model. We will also 
 remark on applying \texttt{GoBlocks} to other models in the $O(N)$ vector series.
 
\subsection{3D Ising Model}
\label{sec:3d_ising}

The 3D Ising model is one of the most extensively studied CFTs in the conformal bootstrap literature. In \cite{Kos:2014bka}, the model was analysed using multiple  correlators involving the lowest-lying 
$\mathbb{Z}_2$-odd and $\mathbb{Z}_2$-even scalars, $\sigma$ and $\epsilon$:
$\langle\sigma\sigma\sigma\sigma\rangle$,
$\langle\sigma\sigma\epsilon\epsilon\rangle$,
$\langle\epsilon\epsilon\epsilon\epsilon\rangle$, and
$\langle\sigma\epsilon\sigma\epsilon\rangle$. 
The system of resulting crossing equations is explicitly given by 
\begin{align}
    \label{eq:final_crossing1}
    0 &= \sum_{\mathcal{O}^+}\lambda_{\sigma\sigma\mathcal{O}}^2 F_{-,\Delta,\ell}^{\sigma\sigma,\sigma\sigma}(u,v) \\
    0 &= \sum_{\mathcal{O}^+}\lambda_{\epsilon\epsilon\mathcal{O}}^2 F_{-,\Delta,\ell}^{\epsilon\epsilon,\epsilon\epsilon}(u,v) \\
    0 &= \sum_{\mathcal{O}^-}\lambda_{\sigma\epsilon\mathcal{O}}^2 F_{-,\Delta,\ell}^{\sigma\epsilon,\epsilon\sigma}(u,v) \\
    \label{eq:final_crossingN}
    0 &= \sum_{\mathcal{O}^+}\lambda_{\sigma\sigma\mathcal{O}}\lambda_{\epsilon\epsilon\mathcal{O}} F_{\mp,\Delta,\ell}^{\sigma\sigma,\epsilon\epsilon}(u,v) 
\pm \sum_{\mathcal{O}^-}(-1)^\ell \lambda_{\sigma\epsilon\mathcal{O}}^2 F_{\mp,\Delta,\ell}^{\epsilon\sigma,\sigma\epsilon}(u,v).
\end{align}

In the linear-functional approach, all five crossing equations can be combined into a single matrix equation which is analysed using semidefinite programming methods. This approach yields highly precise values for the dimensions of the lowest-lying scalars, $\Delta_\sigma = 0.51820(14)$ and $\Delta_\epsilon = 1.4127(11)$. 
Achieving high-precision results requires fine control over the accuracy of the conformal blocks, which \texttt{scalar\_blocks} provides. 

\begin{table}[bt]
\centering
\small
\caption{Spin partition used in the search for the truncated 3D Ising spectrum, consisting
of the 18 stable operators with $\Delta \leq 8$
from~\cite{Simmons_Duffin_2017}. Each $\mathbb{Z}_2$-even operator
contributes three optimisation variables
$(\Delta, \lambda_{\sigma\sigma\mathcal{O}}, \lambda_{\epsilon\epsilon\mathcal{O}})$
and each $\mathbb{Z}_2$-odd operator contributes two,
$(\Delta, \lambda_{\sigma\epsilon\mathcal{O}})$, giving 47 optimisable variables in total.}
\label{tab:spin_partition}

\begin{tabular}{l *{11}{c} *{7}{c}}
\toprule
& \multicolumn{11}{c}{$\mathbb{Z}_2$-even Spectrum} &
\multicolumn{7}{c}{$\mathbb{Z}_2$-odd Spectrum} \\
\cmidrule(lr){2-12}\cmidrule(lr){13-19}
Number      & 1 & 2 & 3 & 4 & 5 & 6 & 7 & 8 & 9 & 10 & 11 &
1 & 2 & 3 & 4 & 5 & 6 & 7 \\
Name   & $\epsilon$ & $\epsilon'$ & & & $T_{\mu\nu}$ & $T'_{\mu\nu}$ & &
$C_{\mu\nu\rho\sigma}$ & & & &
$\sigma$ & $\sigma'$ & & & & & \\
Spin     & 0 & 0 & 0 & 0 & 2 & 2 & 2 & 4 & 4 & 4 & 6 &
0 & 0 & 2 & 2 & 3 & 4 & 5 \\
Num Vars      & 3 & 3 & 3 & 3 & 3 & 3 & 3 & 3 & 3 & 3 & 3 &
2 & 2 & 2 & 2 & 2 & 2 & 2 \\
\bottomrule
\end{tabular}
\end{table}

In contrast, \texttt{GoBlocks} is better suited for situations where such high precision is not essential. This includes the case of truncation methods \cite{Gliozzi:2013ysa,Gliozzi:2014jsa,Gliozzi:2015qsa,Gliozzi:2016cmg,Li:2017ukc,Kantor:2021kbx,Kantor:2021jpz, Kantor:2022epi,Li:2023tic}. These methods typically recast the crossing equations as a nonlinear optimisation problem, aiming to minimise a non-convex loss function over a set of variables subject to constraints. In this section we present results obtained by treating the mixed correlator bootstrap of the 3D Ising model in this nonlinear optimisation context, using \texttt{GoBlocks} for fast online computation of conformal blocks. Our main goal is not to produce new results in the 3D Ising model, but to exhibit the performance of \texttt{GoBlocks} in a realistic application.

Truncation methods require the specification of a list of operators that are included in the optimisation, called the spin partition. In the reported application, our spin partition consisted of the 18 stable
operators with $\Delta \leq 8$ identified in~\cite{Simmons_Duffin_2017}. \cref{tab:spin_partition} lists these operators, grouped by
$\mathbb{Z}_2$ parity and spin. In the mixed-correlator bootstrap, each
$\mathbb{Z}_2$-even operator $\mathcal{O}$ appears in the $\sigma\times\sigma$
and $\epsilon\times\epsilon$ OPEs with independent coefficients
$\lambda_{\sigma\sigma\mathcal{O}}$ and $\lambda_{\epsilon\epsilon\mathcal{O}}$,
contributing three free parameters $(\Delta_\mathcal{O},
\lambda_{\sigma\sigma\mathcal{O}}, \lambda_{\epsilon\epsilon\mathcal{O}})$.
Each $\mathbb{Z}_2$-odd operator appears in the $\sigma\times\epsilon$
OPE, contributing two free parameters
$(\Delta_\mathcal{O}, \lambda_{\sigma\epsilon\mathcal{O}})$ for a total of 47 optimisation variables.
Note that all scaling dimensions, including those of the external
operators $\sigma$ and $\epsilon$ and of the stress tensor $T_{\mu\nu}$,
are treated as free parameters in the optimisation.

The OPE coefficients for this system of equations can be modelled in several ways. We used the following three combinations: $\lambda_{\sigma\sigma\mathcal{O}}$ and $\lambda_{\epsilon\epsilon\mathcal{O}}$ for the $\mathbb{Z}_2$-even spectrum, and $\lambda_{\sigma\epsilon\mathcal{O}}$ for the $\mathbb{Z}_2$-odd spectrum. This approach minimises the number of optimisation variables without introducing extra constraints, while allowing all variables to also take negative values. Other modelling strategies for the OPE coefficients were explored, but performance was similar. The loss function used in the optimisation step was defined as the sum of the vector norms of Eqs.~\eqref{eq:final_crossing1}--\eqref{eq:final_crossingN}, while the ground-truth values against which we compare the optimisation results are taken
from~\cite{Simmons_Duffin_2017}. Constraints were also imposed to ensure the scaling dimensions of operators for each spin were ordered. The OPE equality $\lambda_{\sigma\sigma\epsilon} = \lambda_{\sigma\epsilon\sigma}$ was enforced.

In addition to the \texttt{GoBlocks} parameters discussed in Section~\ref{sec:goblocks}, other block-related parameters influence the accuracy of the optimisation. For the multi-point approach, both the number and placement of sampled points in the complex $z$ plane are important, and na\"ively sampling points near the crossing-symmetric point leads to poor performance. Diagnostics using the known spectrum from Table 2 of \cite{Simmons_Duffin_2017} were used to select an effective set of 100 sampling points, shown in \cref{fig:optimal_z}. With this choice, the total crossing violation was of the order of $10^{-2}$.

As with the multi-point approach, the normalisation of conformal blocks by derivative order $(m,n)$ is crucial for accuracy in the derivative approach. The absolute values of block derivatives can vary widely. Without normalisation by a factor appropriate to the derivative order, the total crossing violation becomes large. In the following experiments, derivative terms were normalised by a factor of  $4^{m+n}\,m!n!$, which ensures that each derivative order contributes comparably to the crossing violation. With this normalisation, the total crossing violation on the known spectrum was again on the order of $10^{-2}$.

\begin{figure}[bt]
    \centering
    \begin{tikzpicture}
\begin{axis}[
    xmin=0.48, xmax=0.62,
    ymin=0.14, ymax=0.76,
    xlabel={Re($z$)},
    ylabel={Im($z$)},
    grid=major,
    width=6cm,
    height=8cm,
]

\foreach \i in {0,...,9}{
    \foreach \j in {0,...,9}{
        \addplot[only marks, mark=*, blue] 
        coordinates {
            ({0.50 + \i*(0.10/9)}, {0.20 + \j*(0.50/9)})
        };
    }
}

\end{axis}
\end{tikzpicture}
    \caption{Sampled points in the complex $z$ plane. This subset of points results in a scalar crossing violation of order $10^{-2}$ when evaluated on the 18 stable operators of the 3D Ising model with dimensions $\Delta \leq 8$, as reported in \cite{Simmons_Duffin_2017}.}
    \label{fig:optimal_z}
\end{figure}

\cref{tab:z_points_performance} presents estimates of the scaling dimensions and OPE coefficients for the lowest-lying scalar operators 
$\sigma$ and $\epsilon$ using the multi-point approach. The non-convex optimisation was carried out with \href{https://github.com/esa/PyGMO2}{PyGMO} \cite{Biscani2020} using the interior-point (IPOPT) algorithm, subject to specified bounds for each search variable. Two sets of bounds were used. Narrow bounds restrict the search to within 25\% of the scaling dimensions and OPE coefficients of all the operators in the spectrum (12.5\% on either side of the true values). Wide bounds allow up to 100\% variation from the true values. In a realistic search scenario, such as that of \cite{Niarchos:2023lot}, a successful strategy involves starting with very wide bounds and reducing them sequentially around the mean values of the CFT data with some tolerance.

\begin{table}[bt]
\centering
\small
\caption{Estimates of the scaling dimensions and OPE coefficients of $\sigma$ and $\epsilon$ in the 3D Ising model using the multi-point approach. Narrow bounds correspond to searching within 25\% of the true values for all data, while wide bounds correspond to 100\%. Both experiments consisted of 100 runs with a population size of $10{,}000$.}
\label{tab:z_points_performance}
\begin{tabular}{c c cc cc}
\toprule
& & \multicolumn{2}{c}{Narrow Bounds} & \multicolumn{2}{c}{Wide Bounds} \\
\cmidrule(lr){3-4} \cmidrule(lr){5-6}
Variable 
& Truth 
& Weighted 
& Unweighted
& Weighted 
& Unweighted \\
\midrule
$\Delta_\sigma$ 
& $0.5181$
& $0.5180 \pm 0.0015$ 
& $0.5186 \pm 0.0035$
& $0.5275 \pm 0.0221$ 
& $0.5332 \pm 0.0245$ \\
$\Delta_\epsilon$ 
& $1.4126$
& $1.4105 \pm 0.0198$ 
& $1.4178 \pm 0.0455$
& $1.4836 \pm 0.3073$ 
& $1.5572 \pm 0.3286$ \\
$\lambda_{\sigma\sigma\epsilon}$ 
& $1.0519$
& $1.0520 \pm 0.0141$ 
& $1.0483 \pm 0.0367$
& $1.0382 \pm 0.2015$ 
& $1.0084 \pm 0.2148$ \\
$\lambda_{\epsilon\epsilon\epsilon}$ 
& $1.5324$
& $1.5244 \pm 0.0263$ 
& $1.5217 \pm 0.0617$
& $1.3994 \pm 0.2703$ 
& $1.3728 \pm 0.3166$ \\
$\lambda_{\sigma\epsilon\sigma}$ 
& $1.0519$
& $1.0516 \pm 0.0133$ 
& $1.0451 \pm 0.0333$
& $0.9967 \pm 0.2300$ 
& $0.9443 \pm 0.2442$ \\
\bottomrule
\end{tabular}
\end{table}

The table reports the unweighted mean and standard deviation (encoding the stochastic spread) over 100 experiments, as well as the mean and standard deviation weighted by the inverse loss returned by the optimiser. Each experiment took approximately three days and required only modest resources: approximately 8\,GB RAM and 6 CPU cores.  In practice, many experiments may be parallelised na\"ively and run synchronously in an array on an HPC cluster.\footnote{All the computations reported here were carried out on the Apocrita HPC cluster at Queen Mary University of London \cite{king_2017_438045}.} To produce OPE coefficients consistent with the conventions in \cite{Simmons_Duffin_2017}, the output of \gbl had to be multiplied by $4^\Delta\,(2\nu)_\ell/(\nu)_\ell$, where $\nu \coloneqq (D-2)/2$.

With narrow search bounds, the estimates for $\Delta_\sigma$, $\lambda_{\sigma\sigma\epsilon}$, and $\lambda_{\sigma\epsilon\sigma}$ are accurate to three decimal places compared to those of \cite{Simmons_Duffin_2017}, while the remaining parameters in \cref{tab:z_points_performance} are accurate to within one or two decimal places. The standard deviations are small, and the true values fall within these intervals. Weighting the results by the inverse loss further improves the estimates, indicating that the loss function effectively models the problem and is sensitive to the $z$-point sampling. When the bounds are widened, the estimates become less accurate, but $\Delta_\sigma$ and $\Delta_\epsilon$ remain within approximately 2\% and 5\% of their true values, demonstrating that the algorithm can reliably explore a large search space.

Table~\ref{tab:derivatives_performance} presents estimates of the conformal data using the derivative approach in \texttt{GoBlocks}. The results are comparable to---though slightly less accurate than---those from the multi-point approach. With narrow bounds, the estimates for $\Delta_\sigma$, $\lambda_{\sigma\sigma\epsilon}$, and $\lambda_{\sigma\epsilon\sigma}$ are accurate to within two decimal places. This experiment employed $n_\text{max}=8$ and $\ell_\text{max}=6$ with a population size of 
$1{,}000$. When the search bounds were widened, the scaling dimensions remained accurate to within 5\% of their true values, using $n_\text{max}=9$ and $\ell_\text{max}=10$. Higher values of these parameters are required for broader search regions. Alternative normalisations of the derivative orders can be used to mitigate bias, but some choices significantly degrade performance. For example, another normalisation resulted in $\Delta_\sigma=0.5147$ and $\Delta_\epsilon=1.3690$. Notably, the unweighted results outperform the weighted ones, suggesting the loss function, which depends on the chosen normalisation, is not yet optimally tuned for this problem.

Similar results have been obtained for the 3D Ising model in the past (see e.g. \cite{Gliozzi:2014jsa}) using truncation methods in the derivative approach. Note, however, that the details of the truncation scheme and the input assumptions in  \cite{Gliozzi:2014jsa} were different. 

\begin{table}[tb]
\centering
\small
\caption{Estimates of the scaling dimensions and OPE coefficients of $\sigma$ and $\epsilon$ in the 3D Ising model using the derivative approach. Both experiments consisted of 100 runs with a population size of $1{,}000$. The narrow bounds used the parameters $n_\text{max}=8$ and $\ell_\text{max}=6$, while wide bounds used $n_\text{max}=9$ and $\ell_\text{max}=10$.}
\label{tab:derivatives_performance}
\begin{tabular}{c c cc cc}
\toprule
& & \multicolumn{2}{c}{Narrow Bounds} & \multicolumn{2}{c}{Wide Bounds} \\
\cmidrule(lr){3-4} \cmidrule(lr){5-6}
Variable 
& Truth 
& Weighted
& Unweighted
& Weighted
& Unweighted \\
\midrule
$\Delta_\sigma$ 
& $0.5181$
& $0.5173 \pm 0.0020$ 
& $0.5188 \pm 0.0041$
& $0.5168 \pm 0.0220$
& $0.5287 \pm 0.0340$ \\
$\Delta_\epsilon$ 
& $1.4126$
& $1.4006 \pm 0.0215$ 
& $1.4167 \pm 0.0437$
& $1.3419 \pm 0.2293$
& $1.4556 \pm 0.3120$ \\
$\lambda_{\sigma\sigma\epsilon}$ 
& $1.0519$
& $1.0569 \pm 0.0163$ 
& $1.0522 \pm 0.0349$
& $1.1212 \pm 0.1302$
& $1.0865 \pm 0.1783$ \\
$\lambda_{\epsilon\epsilon\epsilon}$ 
& $1.5324$
& $1.5131 \pm 0.0379$ 
& $1.5143 \pm 0.0605$
& $1.4334 \pm 0.2495$
& $1.3789 \pm 0.2663$ \\
$\lambda_{\sigma\epsilon\sigma}$ 
& $1.0519$
& $1.0559 \pm 0.0155$ 
& $1.0452 \pm 0.0319$
& $1.0938 \pm 0.1653$
& $1.0116 \pm 0.2265$ \\
\bottomrule
\end{tabular}
\end{table}

\subsection{Scaling Up in the $O(N)$ Model Series}

The multi-correlator numerical bootstrap in the linear-functional approach has been extended beyond the Ising model in the $O(N)$ vector model series (see e.g. \cite{Chester:2019ifh,Chester:2020iyt}).
Higher values of $N$ introduce a significant increase in compute requirements. For example, the three main searches in the $O(3)$ bootstrap of \cite{Chester:2020iyt} together required approximately 3  
million CPU-hours. 
It is interesting to ask how things scale up in a truncation scheme. In this section we discuss the computational complexity involved when extending the use of \gbl to the $O(2)$ and $O(3)$ vector models.

Recall that in the derivative approach to \texttt{GoBlocks}, there are three major computational steps: recursion, evaluation, and conversion (see \cref{fig:deriv_flowchart}). 
The recursion step is fast ($\sim\!50$\,ms for \texttt{n\_max}~$= 8$) and its runtime is independent of $\Delta_{ij}$ and $\Delta_{kl}$. The conversion step dominates for large \texttt{n\_max} on the first call, but its cost is amortised through caching. Alternatively, \gbl supports independent cache building, preventing potentially memory-intensive processes being called at runtime, and thus allowing cluster resources to be tuned without exceeding allocation limits on the first call. The evaluation and conversion steps scale linearly with the number of exchanged operators, but are much less computationally expensive. Conversion to derivatives of $F_+$ must be computed separately from derivatives of $F_-$, as they involve different derivative orders $(m,n)$. The total computation time for a single \texttt{GoBlocks} call is the sum of these steps. Further optimisation is possible across multiple calls, such as during an optimisation over different operators and representations. From  \cref{eq:scalar_blocks_output}, $g$ depends only on the differences $\Delta_{ij}$ and $\Delta_{kl}$, not on the dimensions of the individual external operators. This allows us to reuse the recursion results when these differences coincide, reducing the total number of recurse steps required. The number of evaluate and convert steps remains unchanged, as they depend on the individual operator dimensions and spins.

\begin{table}[bt]
\centering
\small
\caption{The number of recurse, evaluate, and convert steps needed for the $O(2)$ model by representation. Assuming the total runtime is dominated by the recurse step, the best case total time per optimiser step is $\sim 0.5$ seconds. Recurse depends only on the differences of the external dimensions $\Delta_{ij}$ and $\Delta_{kl}$, meaning some computations can be reused; na\"ive and best refer to the number of steps both without and with reuse, respectively.}
\label{tab:o2_model_computationl_steps}
\begin{tabular}{rccccccccc}
\toprule
\multicolumn{1}{c}{} & \multicolumn{7}{c}{\textbf{Representations}} \\
\cmidrule(lr){2-8}
\textbf{Steps} & $(\ell^+,0^+)$ & $(\ell^-, 0^-)$ & $(\ell^\pm, 1)$ & $(\ell^+, 2)$ & $(\ell^-, 2)$ & $(\ell^\pm, 3)$ & $(\ell^+, 4)$ & \textbf{Na\"ive} & \textbf{Best} \\
\midrule
Recurse  & 6 & 3 & 6 & 4 & 2 & 2 & 1 & 24 & 10 \\
Evaluate & 6 & 3 & 6 & 4 & 2 & 2 & 1 & 24 & 24 \\
Convert  & 11 & 6 & 10 & 7 & 3 & 4 & 2 & 43 & 43 \\
\bottomrule
\end{tabular}
\end{table}

The structure of the crossing equations directly determines the number of computational steps per optimiser iteration. For each distinct pair of external operator differences $\Delta_{ij},\Delta_{kl}$ where blocks are evaluated, add one recurse and one evaluate step. For every unique combination of $\Delta_{ij},\Delta_{kl}$ and $F_\pm$, add a convert step. 

    In the Ising model, \cref{eq:final_crossing1} to \cref{eq:final_crossingN} yield five unique $\Delta_{ij},\Delta_{kl}$ permutations: three from the first three equations and two from the last. Only three recurse steps are needed, since $\langle \sigma\sigma\sigma\sigma\rangle$, $\langle\epsilon\epsilon\epsilon\epsilon\rangle$, and $\langle\sigma\sigma\epsilon\epsilon\rangle$ correspond to $\Delta_{ij} = 0 = \Delta_{kl}$ and hence the same conformal block $g$. There are seven unique $\Delta_{ij},\Delta_{kl}$ and $F_\pm$ combinations: three from the first three equations and four from the last. The total runtime per optimiser step is 
$3t_\text{recurse} +5t_\text{evaluate} + 7t_\text{convert}$. If the recurse step dominates and its runtime is given by \cref{fig:average_runtime_vs_accuracy}, each step takes $3\times 0.035 = 0.105$\,s.

The same estimates can be applied to the $O(2)$ and 
$O(3)$ models. In the $O(2)$ case, operators fall into irreducible representations (irreps) of $O(2)\cong U(1)\times \mathbb{Z}_2$. Beyond the trivial $\mathbf{0}^+$ and sign $\mathbf{0}^-$ representations, there is an infinite family of two-dimensional irreps labelled by $q$, spanned by states with $U(1)$ charge $\pm q$ exchanged by $\mathbb{Z}_2$. In general, irreps are labelled $(\ell^\pm, \textbf{q}^\pm)$, where $\ell^\pm$ denotes spin and spatial parity, and $\mathbf{q}^\pm$ denotes the $q$ irrep and its $\mathbb{Z}_2$ parity. There are 22 crossing equations, as given in Equation~(17) of \cite{Chester:2019ifh}. The number of computational steps can be directly read off from the $\vec{V}$ vectors for each irrep in that equation. \cref{tab:o2_model_computationl_steps} summarises the steps per irrep in the various crossing channels. In total, there are 24 recurse and evaluate steps, and 43 convert steps. The 24 recurse steps reduce to 10 due to redundancies. Assuming the recurse step dominates, the total runtime is $10 \times 0.035 = 0.35$\,s.

For the $O(3)$, or Heisenberg, model, irreducible representations are also labelled by $\mathbf{q}^\pm$. There are 28 crossing equations given in Eq.~(14) of \cite{Chester:2020iyt}, with $\vec{V}$ vectors provided in a separate notebook. \cref{tab:o3_model_computationl_steps} lists the number of processing steps per representation. The minimal number of recurse steps required is 11, only one more than in the $O(2)$ model, yielding a best-case runtime of $0.55$\,s. The computational complexity of the $O(3)$ model is comparable to that of $O(2)$, particularly when redundancies in $g$ are taken into account.

\begin{table}[bt]
\centering
\scriptsize
\caption{The number of recurse, evaluate, and convert steps needed for the $O(3)$ model broken out by representation. Assuming the total runtime is dominated by the recurse step, the total time per optimiser step is of the order of $0.55$ seconds.}
\label{tab:o3_model_computationl_steps}
\setlength{\tabcolsep}{5pt}
\begin{tabular}{r ccccccccc cc}
\toprule
\multicolumn{1}{c}{} & \multicolumn{9}{c}{\textbf{Representations}} \\
\cmidrule(lr){2-10}
\textbf{Steps} & $(\ell^+, 0^+)$ & $(\ell^-, 1^+)$ & $(\ell^\pm, 1^-)$ & $(\ell^+, 2^+)$ & $(\ell^-, 2^+)$ & $(\ell^\pm, 2^-)$ & $(\ell^-, 3^+)$ & $(\ell^\pm, 3^-)$ & $(\ell^+, 4^+)$ & \textbf{Na\"ive} & \textbf{Best} \\
\midrule
Recurse  & 6 & 3 & 6 & 7 & 2 & 2 & 1 & 2 & 1 & 30 & 11 \\
Evaluate & 6 & 3 & 6 & 7 & 2 & 2 & 1 & 2 & 1 & 30 & 30 \\
Convert  & 11 & 6 & 10 & 12 & 3 & 4 & 2 & 4 & 2 & 54 & 54 \\
\bottomrule
\end{tabular}
\end{table}

The following points warrant special mention. First, these estimates do not account for scaling with the number of operators in the truncation. Both the evaluate and convert steps scale with this number and should be included in realistic runtime estimates. In the Ising model, most computation time is spent in the recurse step. Second, this analysis applies only to the derivative approach in \texttt{GoBlocks}. In the points approach, there is no convert step, but recurse and evaluate must be performed twice, once for $g(u,v)$ and once for $g(v, u)$, for each $z$ point. Despite this, the points approach is significantly faster than the derivatives approach, so actual runtimes can be much lower, even when accounting for the additional recursions.

\section{Outlook}
\label{sec:outlook}

In this work we introduced \texttt{GoBlocks}, a fast, parallel, and flexible conformal block generator supporting both multi-point and derivative-based bootstrap strategies. The \gbl package is designed for scenarios where computational speed and moderate accuracy are critical, but ultra-high precision is not essential. In such contexts, the widely used \sbl package can become a computational bottleneck.

The first part of this study benchmarked \gbl against existing methods in terms of both speed and accuracy. The multi-point approach in \gbl achieves agreement with \sbl at the crossing-symmetric point to within an average percent error as low as
 $2.5\times 10^{-8}$ and agreement with exact analytic expressions along the $\bar z = z$ diagonal at median $5\times 10^{-9}$.
 Overall, for block accuracies below $10^{-4}\%$, and likely even higher, \gbl is able to attain  up to a five-fold speed increase over the  \texttt{scalar\_blocks} benchmark.

In the second part, \gbl was applied in directly solving for the CFT data of the 3D Ising model with hard truncation, formulated as a non-convex optimisation problem. With narrow parameter bounds around the state-of-the-art numerical solution, the optimiser recovered most low-lying CFT data to within three decimal places. For wider bounds, the optimiser achieved results within 2--5\% of the true values, demonstrating the robustness of the approach. The computational scaling was further analysed by quantifying the number of steps required per optimiser iteration in the $O(2)$ and $O(3)$ models. The algorithm exhibits favourable scaling, with both models requiring only about 0.5 seconds per step using the derivative approach. The multi-point approach is substantially faster, offering even lower runtimes with similar accuracy.

The multi-point and derivative approaches to conformal-block evaluation in \texttt{GoBlocks} offer complementary trade-offs. The multi-point approach is five to ten times faster, owing to the absence of a convert step, and in the 3D Ising bootstrap it recovered $\Delta_\sigma, \Delta_\epsilon$ and the leading OPE coefficients to three decimal places with narrow search bounds, compared to two decimal places for the derivative approach under comparable conditions. However, each method introduces its own source of tuning: the multi-point approach requires a choice of sampling points in the $z$-plane to ensure rapid OPE convergence~\cite{CastedoEcheverri:2016swn}, whereas the derivative approach is sensitive to the normalisation of derivative orders, with suboptimal choices significantly degrading performance.

These tuning choices have a counterpart in interpolation-based methods, where accuracy hinges on how the blocks are sampled. In~\cite{Chang:2025mwt}, the polynomial $P$ of \cref{eq:scalar_blocks_output} is approximated using optimal density nodes for the weight $\mu(\Delta) \coloneqq r_0^\Delta / Q_{\kappa,\ell}(\Delta)$, which improves accuracy against the true blocks at low to intermediate scaling dimensions. The technique requires only the ability to evaluate $P$ at a chosen set of scaling dimensions. While \gbl does not expose this quantity directly, it could be modified to return it, enabling an interpolated variant to be benchmarked against \texttt{scalar\_blocks}. The same work also treats the optimal estimation of functionals acting on the blocks, which arise in the dual formulation of the bootstrap. There, interpolating the analytically known optimal functional of the spin-two gap problem, or dispersive functional, again yields improved estimation. This is most relevant to the dual bootstrap, whereas our applications target the primal bootstrap, where the closest analogues are the choice of sampling points and the normalisation of derivative orders. A systematic study of how these parameters affect the recovery of CFT data across models would be a natural next step.

The difficulties discussed above are well-known inherent issues of the hard truncation approach, which we expect to be mitigated in the future by more systematic, flexible and reliable high-dimensional search strategies in the conformal bootstrap, aiming not only to constrain low-lying CFT data, but also to reconstruct full correlation functions. Conformal-block evaluation tools, like \sbl and \texttt{GoBlocks}, will be useful in such developments. In this context, it would be interesting to explore novel methodologies that combine the speed of \gbl with the high-precision of \sbl (for instance, in potential variants of search algorithms like the Navigator function \cite{Reehorst:2021ykw}). It would also be interesting to explore variants of \gbl itself beyond the four-point scalar bootstrap, e.g.\ for higher-spin \cite{Erramilli:2020rlr} or higher-point \cite{Goncalves:2019znr,Poland:2021xjs} conformal blocks. 

\section{Data Availability}

Along with this paper, we have made two packages publicly available:

\begin{enumerate}
    \item \texttt{GoBlocks}: Available at \href{https://github.com/xand-stapleton/goblocks}{\texttt{https://github.com/xand-stapleton/goblocks}}.
    \item \texttt{BootSTOP-multi-correlator}: The multi-objective implementation of the bootstrap stochastic optimiser, \texttt{BootSTOP} \cite{Kantor:2021jpz, Kantor:2022epi, Kantor:2021kbx, Niarchos:2023lot}. Available at \url{https://github.com/jchryssanthacopoulos/BootSTOP-multi-correlator}.
\end{enumerate}

\section*{Acknowledgments}
The authors would like to thank P. Richmond for collaboration at early stages of this work and S.~Rychkov for correspondence on the implementation of the recursive algorithm. CP was partially supported by the Science and Technology Facilities Council (STFC) Consolidated Grant ST/X00063X/1 “Amplitudes, Strings \& Duality.” AGS acknowledges support from Pierre Andurand. This research utilised the Apocrita HPC facility, supported by QMUL Research-IT \cite{king_2017_438045}.

\begin{appendix}
\crefalias{section}{appendix}

\section{Interpolation Schemes for Conformal Blocks}
\label{app:multilinear_polynomial_interpolation}

A fast way of generating block derivatives at the crossing-symmetric point involves pre-computing block values with \sbl on a fixed grid of scaling dimensions, followed by multilinear interpolation to estimate values at arbitrary points in the space of scaling dimensions. This method offers high speed and straightforward implementation, though accuracy may degrade in regions where the blocks exhibit significant variation or lack smoothness. One practical realisation employs a $k$-$d$ tree to identify the $k$ nearest neighbours to a target point. The corresponding $F$-block values, as defined in \cref{eq:F_derivatives}, are then combined using weights inversely proportional to their distances from the query location.

To assess the accuracy of multilinear interpolation, a reference set of 
$F$-block values was generated on a new grid of scaling dimensions with  $100{,}000$ points. For simplicity, this grid was restricted to spin $\ell = 0$. The interpolation error was computed by comparing the interpolated $F$ values to these reference values. \cref{tab:multilinear_interpolation_performane} reports percentiles of the average component error for various choices of the number of neighbours 
$k$. The average component error is defined as the mean relative error across all components of the $F$-block derivatives. At the 25th percentile, errors remain moderate, typically in the range of 14--30\%. At the 90th percentile, errors increase substantially, reaching 200--300\%. The lowest overall errors are observed for $k=2$ neighbours. Higher errors are expected for larger spin values.

The results show that, although multilinear interpolation is computationally efficient, it does not provide sufficient accuracy for most applications. The precomputed block grid was sampled with a spacing of $0.1$ in the scaling dimensions, so the final optimisation results are limited to an accuracy of $\delta\Delta=0.1$. Increasing the grid resolution can improve accuracy, but this requires significantly more memory to store and access the blocks, which constrains the practical resolution achievable with this approach.

\begin{table}[t]
\centering
\small
\caption{Percent errors of block derivatives for multilinear interpolation as a function of the number of neighbours $k$.}
\label{tab:multilinear_interpolation_performane}
\begin{tabular}{ccccc}
\toprule
\multicolumn{1}{c}{\textbf{Parameters}} & \multicolumn{4}{c}{\textbf{Error (\%)}} \\
\cmidrule(lr){1-1} \cmidrule(lr){2-5}
$k$ & 25\% & 50\% & 75\% & 90\% \\
\midrule
1  & 19.32 & 42.28 & 91.97 & 195.26 \\
2  & 14.34 & 32.67 & 71.61 & 167.86 \\
4  & 14.22 & 32.12 & 75.61 & 188.50 \\
8  & 20.28 & 44.84 & 105.06 & 257.85 \\
16 & 30.39 & 66.65 & 154.56 & 374.81 \\
\bottomrule
\end{tabular}
\end{table}

Polynomial interpolation is an alternative to multilinear interpolation. In this approach, the $k$ nearest neighbours to a query point $x$ are identified and used to construct a design matrix $X_\text{poly}$, where each column represents a polynomial feature of degree up to $n$ and each row corresponds to a neighbour. The associated block values are collected in the vector $y$. The interpolation coefficients $\beta$ are obtained by solving the ridge regression problem
\begin{equation}
\left(X_\text{poly}^T X_\text{poly} + \lambda I\right)\beta = X_\text{poly}^T y
\end{equation}
for a given regularisation parameter $\lambda$. The interpolated block value at the query point is then given by $\vec{x}_\text{poly}\cdot\beta$, where $\vec{x}_\text{poly}$ is the vector of polynomial features evaluated at $x$.

\cref{tab:polynomial_interpolation_performane} presents error statistics for the five best-performing parameter choices, ranked by the 90th percentile error. Polynomial interpolation yields lower errors than multilinear interpolation, though at the cost of increased runtime. Despite this improvement, the error levels remain too high for use in precision bootstrap analyses. These limitations motivate the development of block representations that can be evaluated directly at arbitrary query points, enabling both speed and accuracy without relying on interpolation.

\begin{table}[t]
\centering
\small
\caption{Percent errors of block derivatives for polynomial interpolation as a function of the number of neighbours $k$, degree of the polynomial $n$, and regularisation $\lambda$.}
\label{tab:polynomial_interpolation_performane}
\begin{tabular}{ccc cccc}
\toprule
\multicolumn{3}{c}{\textbf{Parameters}} & \multicolumn{4}{c}{\textbf{Error (\%)}} \\
\cmidrule(lr){1-3} \cmidrule(lr){4-7}
$k$ & $n$ & $\lambda$ & 25\% & 50\% & 75\% & 90\% \\
\midrule
16 & 2 & 0.001 & 9.55 & 21.38 & 51.71 & 136.40 \\
8  & 2 & 0.001 & 11.09 & 25.12 & 59.17 & 144.31 \\
8  & 3 & 0.100 & 10.04 & 24.00 & 59.63 & 150.38 \\
16 & 3 & 0.100 & 9.83 & 22.12 & 56.04 & 151.84 \\
2  & 2 & 0.001 & 15.59 & 32.89 & 68.88 & 155.14 \\
\bottomrule
\end{tabular}
\end{table}

A more principled approach to interpolation is developed in~\cite{Chang:2025mwt}, where the nodes are placed not on a uniform grid but at the optimal density for the weight $\mu(\Delta) \coloneqq r_0^\Delta / Q_{\kappa,\ell}(\Delta)$, improving accuracy against the true blocks at low to intermediate scaling dimensions. This targets directly the failure mode seen above, where a fixed grid spacing both caps the achievable resolution and forces errors to grow in regions of rapid block variation. Adapting these node-placement strategies to the block representations used here would be a natural way to recover the speed of interpolation without the accuracy penalties reported above.

\section{Algorithmic Flow of GoBlocks}
\label{app:schematic_overview}

The package \gbl supports both multi-point and derivative approaches. \cref{fig:multipoint_flowchart} shows the algorithmic flow of the multi-point approach, while \cref{fig:deriv_flowchart} depicts the flow of the derivative approach. In the multi-point approach, the recursion step computes $r$-series coefficients sequentially (each order depends on the previous ones), but the evaluation and $F_\pm$ construction steps are parallelised across $z$-points. In the derivative approach, each successive derivative order depends on all lower orders, further limiting parallelisation.

\begin{figure}
    \centering
    \resizebox{!}{0.85\textheight}{%
\begin{tikzpicture}[
  box/.style  = {rectangle, draw=black!55, rounded corners=4pt,
                 fill=#1, text width=7cm, minimum width=9cm,
                 align=center, inner sep=7pt, font=\small},
  box/.default = white,
  arr/.style  = {->, >=Stealth, thick, black!65},
  node distance = 8mm,
]

\node[box=blue!8] (inputs)
{\textbf{Inputs:} External dims. $(\Delta_{ij},\Delta_{kl})$;
evaluation points $\{z_1,\dots,z_N\}$; spacetime dim.\ $D$;\\
operator data $(\Delta_a,\ell_a)$; block types $(F_+,F_-)$};

\node[box=teal!8, below=18mm of inputs] (setup)
{\textbf{Setup:} compute parameters,\\
convert $z_i\to(r,\eta)$ and $1{-}z_i\to(r_1,\eta_1)$};

\node[box=teal!8, below=10mm of setup] (poles)
{\textbf{Determine recursion poles:}\\
compute pole positions, coefficients, and residues for each spin};

\node[box=teal!8, below=10mm of poles] (seed)
{\textbf{Seed values:}\\
compute Taylor coefficients of $\tilde h^{\Delta_{ij},\Delta_{kl}}_{\ell}(r)$
around $r=0$};

\node[box=teal!8, below=10mm of seed] (iter)
{\textbf{$r$-series recursion:}\\
{Compute Taylor coefficients of $h$ order by order.
}};

\node[box=red!8, below=14mm of iter] (eval)
{\textbf{Evaluate $g$ blocks:}\\
for each $(\Delta_a,\ell_a)$, compute $g$\\
at both $(r,\eta)$ and $(r_1,\eta_1)$};

\node[box=orange!10, below=18mm of eval] (fblock)
{\textbf{Construct $F_\pm$ blocks:}\\
$F_\pm = v^{\bar\Delta_{jk}}\,g(u,v)
         \pm u^{\bar\Delta_{jk}}\,g(v,u)$};

\node[box=blue!8, below=18mm of fblock] (output)
{\textbf{Output:}
$F_{\pm,\Delta_a,\ell_a}(z_i,\bar z_i)$
for all $z_i$};

\draw[arr] (inputs) -- (setup);
\draw[arr] (setup) -- (poles);
\draw[arr] (poles) -- (seed);
\draw[arr] (seed) -- (iter);
\draw[arr] (iter) -- (eval);
\draw[arr] (eval) -- (fblock);
\draw[arr] (fblock) -- (output);

\begin{pgfonlayer}{background}

\node[draw=teal!50, fill=teal!5, rounded corners=6pt, line width=0.8pt,
      fit=(setup)(poles)(seed)(iter),
      inner xsep=14pt, inner ysep=20pt] (p1) {};

\node[draw=red!45, fill=red!5, rounded corners=6pt, line width=0.8pt,
      fit=(eval)(p1.west |- eval)(p1.east |- eval),
      inner xsep=0pt, inner ysep=20pt] (p2) {};

\node[draw=orange!45, fill=orange!6, rounded corners=6pt, line width=0.8pt,
      fit=(fblock)(p1.west |- fblock)(p1.east |- fblock),
      inner xsep=0pt, inner ysep=20pt] (p3) {};

\end{pgfonlayer}

\node[font=\footnotesize\bfseries\color{teal!65!black},
      fill=teal!5, draw=teal!50, rounded corners=2pt,
      inner sep=3pt, anchor=north west]
  at ([xshift=4pt,yshift=-4pt] p1.north west)
  {Phase 1 --- Recursion};

\node[font=\footnotesize\bfseries\color{red!65!black},
      fill=red!5, draw=red!45, rounded corners=2pt,
      inner sep=3pt, anchor=north west]
  at ([xshift=4pt,yshift=-4pt] p2.north west)
  {Phase 2 --- Block evaluation};

\node[font=\footnotesize\bfseries\color{orange!70!black},
      fill=orange!6, draw=orange!45, rounded corners=2pt,
      inner sep=3pt, anchor=north west]
  at ([xshift=4pt,yshift=-4pt] p3.north west)
  {Phase 3 --- $F_\pm$ construction};

\end{tikzpicture}%
    }
    \caption{Schematic overview of multi-point block evaluation steps. 
    }
    \label{fig:multipoint_flowchart}
\end{figure}

\begin{figure}
    \centering
    \resizebox{0.85\textwidth}{!}{\begin{tikzpicture}[
  box/.style  = {rectangle, draw=black!55, rounded corners=4pt,
                 fill=#1, text width=7cm, minimum width=9cm,
                 align=center, inner sep=7pt, font=\small},
  box/.default = white,
  dec/.style  = {diamond, draw=black!60, fill=teal!12,
                 text width=2.4cm, align=center,
                 aspect=2.4, inner sep=2pt, font=\small},
  arr/.style  = {->, >=Stealth, thick, black!65},
  node distance = 8mm,
]

\node[box=blue!8] (in1)
{\textbf{Recursion inputs:}\\ External dims. $(\Delta_{ij},\Delta_{kl})$;\\ Max. deriv. order $n_\text{max}$};

\node[box=teal!8, below=18mm of in1] (setup)
{\textbf{Setup:} compute parameters, $\tilde h^{\Delta_{ij}, \Delta_{kl}}_{\ell}(r, \eta)$};

\node[box=teal!8, below=10mm of setup] (poles)
{\textbf{Determine recursion poles:}\\
compute pole positions, coefficients, and residues for each spin};

\node[box=teal!8, below=10mm of poles] (seed)
{\textbf{Seed derivatives:}\\
evaluate derivatives of seed functions at $(r_*,\eta_*)$};

\node[box=teal!8, below=10mm of seed] (iter)
{\textbf{Recursive update:}\\
Calculate successive $dh_\text{new}$ by applying recursion matrices and pole contributions};

\node[dec, below=10mm of iter] (conv)
{Converged?};

\node[dec, right=20mm of conv] (maxit)
{\texttt{max\_iter}\\hit?};

\node[box=red!8, below=25mm of conv] (eval)
{\textbf{Evaluate block derivatives:}\\
compute derivatives for each operator $(\Delta_a,\ell_a)$};

\node[box=blue!8, left=18mm of eval, text width=4.5cm, minimum width=5cm, font=\footnotesize] (in2)
{\textbf{Operator data:} $(\Delta_a,\ell_a)$\\
spacetime dim.\ $D$, block types $(F_+,F_-)$};


\node[box=orange!10, below=25mm of eval] (coords)
{\textbf{Coordinate transformation:}\\
convert derivatives from $(r,\eta)$ to $(z,\bar z)$};

\node[box=orange!10, below=of coords] (fblock)
{\textbf{Construct $F_\pm$ blocks:}\\
combine derivatives to form even ($F_+$) or odd ($F_-$) blocks};

\node[box=blue!8, below=18mm of fblock] (output)
{\textbf{Output:} derivatives
$\{\partial_z^m\partial_{\bar z}^n F_{\pm_,\Delta_a,\ell_a}\}$};


\draw[arr] (in1) -- (setup);

\draw[arr] (in2.east) |- (eval.west);


\draw[arr] (setup) -- (poles);
\draw[arr] (poles) -- (seed);
\draw[arr] (seed) -- (iter);
\draw[arr] (iter) -- (conv);

\draw[arr] (conv.south) -- node[right,font=\footnotesize]{Yes} (eval.north);
\draw[arr] (conv.east) -- node[above,font=\footnotesize]{No} (maxit.west);

\draw[arr] (maxit.north) |- node[above,font=\footnotesize]{No} (iter.east);

\draw[arr] (maxit.south) |- node[right,font=\footnotesize]{Yes} (eval.east);

\draw[arr] (eval) -- (coords);
\draw[arr] (coords) -- (fblock);
\draw[arr] (fblock) -- (output);

\begin{pgfonlayer}{background}

\node[draw=teal!50, fill=teal!5, rounded corners=6pt, line width=0.8pt,
      fit=(setup)(poles)(seed)(iter)(conv)(maxit),
      inner xsep=14pt, inner ysep=20pt] (p1) {};

\node[draw=red!45, fill=red!5, rounded corners=6pt, line width=0.8pt,
      fit=(eval)(p1.west |- eval)(p1.east |- eval),
      inner xsep=0pt, inner ysep=20pt] (p2) {};

\node[draw=orange!45, fill=orange!6, rounded corners=6pt, line width=0.8pt,
      fit=(coords)(fblock)(p1.west |- coords)(p1.east |- fblock),
      inner xsep=0pt, inner ysep=20pt] (p3) {};

\end{pgfonlayer}

\node[font=\footnotesize\bfseries\color{teal!65!black},
      fill=teal!5, draw=teal!50, rounded corners=2pt,
      inner sep=3pt, anchor=north west]
  at ([xshift=4pt,yshift=-4pt] p1.north west)
  {Phase 1 — Recursion};

\node[font=\footnotesize\bfseries\color{red!65!black},
      fill=red!5, draw=red!45, rounded corners=2pt,
      inner sep=3pt, anchor=north west]
  at ([xshift=4pt,yshift=-4pt] p2.north west)
  {Phase 2 — Block evaluation};

\node[font=\footnotesize\bfseries\color{orange!70!black},
      fill=orange!6, draw=orange!45, rounded corners=2pt,
      inner sep=3pt, anchor=north west]
  at ([xshift=4pt,yshift=-4pt] p3.north west)
  {Phase 3 — Coordinate conversion and $F^\pm$ construction};

\end{tikzpicture}}
    \caption{Schematic overview of block derivative evaluation steps.}
    \label{fig:deriv_flowchart}
\end{figure}

\section{Derivation of a Recursive Relation for Block Derivatives}
\label{app:recursive_deriv}

We wish to find a recursive relation for derivatives of the conformal
blocks $\partial_r^m \partial_\eta^n h^{\Delta_{ij},
\Delta_{kl}}_\ell(r, \eta)$ given the equation
\begin{equation}
  \label{eqn:big_daddy_recursive_2}
  h_{\Delta, \ell}^{\Delta_{ij}, \Delta_{kl}}  (r, \eta) =
  \tilde{h}_{\ell}^{\Delta_{ij}, \Delta_{kl}}  (r, \eta) + \sum_{i}
  \frac{c_i^{\Delta_{ij}, \Delta_{kl}}}{\Delta-\Delta_i} r^{n_i}
  h_{\Delta_i + n_i, \ell_i}^{\Delta_{ij}, \Delta_{kl}}  (r, \eta),
\end{equation}
where as in \cite{Kos:2014bka}, we define
\begin{equation}
  \tilde h^{\Delta_{ij}, \Delta_{kl}}_\ell(r, \eta) =
  \frac{\ell!}{(2\nu)_\ell} \frac{(-1)^\ell C^\nu_\ell (\eta)(1+r^2
  + 2r\eta)^{\alpha} (1+r^2 - 2r\eta)^{\beta}}{(1-r^2)^\nu } ,
\end{equation}
with $C_\ell^\nu(\eta)$ denoting the ordinary Gegenbauer polynomial, with $\nu \coloneqq (D-2)/2$, and
\begin{equation}
  \alpha \coloneqq -\frac{1}{2} (1 + \Delta_{ij} - \Delta_{kl}),\quad\beta \coloneqq
  -\frac{1}{2} (1 - \Delta_{ij} + \Delta_{kl}).
\end{equation}
The pole coefficients $c_i$ are defined in a table in
\cite{Kos:2014bka}, so we refer the reader there. The main
tool in this calculation will be the Leibniz theorem, which states:
\begin{equation}
  \label{liebniz_multivar}
  \partial_r^m \partial_\eta^n f(r, \eta) g(r, \eta) = \sum_{i=0}^{m}
  \sum_{j=0}^{n} \binom{m}{i} \binom{n}{j} \partial_r^i
  \partial_\eta^j f(r, \eta) \partial_r^{m-i} \partial_\eta^{n-j} g(r, \eta).
\end{equation}

Applying \cref{liebniz_multivar} to the second term in
\cref{eqn:big_daddy_recursive_2}, we obtain
\begin{multline}
  \partial_r^m \partial_\eta^n h_{\Delta, \ell}^{\Delta_{ij},
  \Delta_{kl}}  (r, \eta) = \partial_r^m \partial_\eta^n
  \tilde{h}_{\ell}^{\Delta_{ij}, \Delta_{kl}}  (r, \eta)\\
  + \sum_{i} \frac{c_i^{\Delta_{ij}, \Delta_{kl}}}{\Delta-\Delta_i}
  \sum_{s=0}^m \sum_{t=0}^n \binom{m}{s} \binom{n}{t}
  \bigg[\partial_r^{m-s} \partial_\eta^{n-t} (r^{n_i}) \cdot
    \partial_r^s \partial_\eta^t (h_{\Delta_i + n_i, \ell_i}^{\Delta_{ij},
  \Delta_{kl}}  (r, \eta))\bigg].
\end{multline}
The sum over $\eta$ collapses since  $\partial_r^{m-s}
\partial_\eta^{n-t} r^{n_i} = (n_i)_{(m-s)}r^{n_i-m+s}\delta_{n, t}$ where $(a)_{(b)}$ denotes the falling factorial. Thus,
\begin{multline}
  \label{eqn:final_recursive_rel}
  \partial_r^m \partial_\eta^n h_{\Delta, \ell}^{\Delta_{ij},
  \Delta_{kl}}  (r, \eta) = \partial_r^m \partial_\eta^n
  \tilde{h}_{\ell}^{\Delta_{ij}, \Delta_{kl}}  (r, \eta)\\
  + \sum_{i} \frac{c_i^{\Delta_{ij}, \Delta_{kl}}}{\Delta-\Delta_i}
  \sum_{s=0}^m \binom{m}{s} \bigg[(n_i)_{(m-s)}r^{n_i-m+s} \cdot
    \partial_r^s \partial_\eta^n (h_{\Delta_i + n_i, {\ell_i}}^{\Delta_{ij},
  \Delta_{kl}}  (r, \eta))\bigg].
\end{multline}

In order to find the full derivative blocks, we first tackle
$\partial_r^m \partial_\eta^n \tilde h^{\Delta_{ij},
\Delta_{kl}}_\ell(r, \eta)$. In order to use \cref{liebniz_multivar},
we let $f(r, \eta)\coloneqq f_1(r, \eta) f_2^\alpha (r, \eta)f_3^\beta(r,
\eta)$, and $g(\eta) \coloneqq C_\ell^\nu(\eta)$, where
\begin{equation}
  f_1 \coloneqq (1-r^2)^{-\nu}, \qquad f_2 \coloneqq 1+r^2 + 2r\eta, \qquad f_3\coloneqq 1+r^2 - 2r\eta.
\end{equation}
In order to calculate the $\partial_r^m \partial_\eta^n f(r, \eta)$
part, we need to once again apply  the Leibniz rule:
\begin{equation}
  \partial_r^i \partial_\eta^j f(r, \eta) =
  \sum_{\substack{i_1+i_2+i_3=i\\j_1+j_2+j_3=j}}
  \frac{i!}{i_1!i_2!i_3!} \frac{j!}{j_1! j_2! j_3!} (\partial_r^{i_1}
  \partial_\eta^{j_1} f_1(r))(\partial_r^{i_2} \partial_\eta^{j_2}
  f_2^\alpha(r, \eta)) (\partial_r^{i_3} \partial_\eta^{j_3}
  f_3^\beta(r, \eta)).
\end{equation}
Notice
\begin{align}
  \partial_r^i \partial^j_\eta f_1(r) \eqqcolon \phi_1^{(i,j)}(r) &=
  \begin{cases}
    \partial_r^i f_1(r) &\quad \text{if } j = 0, \\
    0 &\quad \text{otherwise.}
  \end{cases}
\end{align}
This equation is independent of the external scaling dimensions and is solely dependent on the spacetime dimension $D$.
Since $\nu$ need not be an integer, one must employ the Fa\`a di Bruno formula. Let
\begin{equation}
h(u) = u^{-\nu}, \qquad u(r) = 1 - r^2,
\end{equation}
such that $\phi_1^{i, 0}(r)$ may be written as
\begin{equation}
\frac{\partial^i}{\partial r^i}(1-r^2)^{-\nu}
=
\frac{\partial^i}{\partial r^i} h(u(r)).
\end{equation}
Using Fa\`a di Bruno's formula,
\begin{equation}
\frac{\partial^i}{\partial r^i} h(u(r))
=
\sum_{k=0}^{\lfloor i/2\rfloor}
h^{(i-k)}(u)
\frac{i!}{k!(i-2k)!}
\left(\frac{u''}{2}\right)^k
(u')^{i-2k}.
\end{equation}
The derivatives of $h$ are
\begin{equation}
h^{(m)}(u) = (-1)^m (\nu)_m \, u^{-\nu-m},
\end{equation}
leading to
\begin{align*}
\frac{\partial^i}{\partial r^i}(1-r^2)^{-\nu}
&=
\sum_{k=0}^{\lfloor i/2\rfloor}
(-1)^{i-k}(\nu)_{i-k}(1-r^2)^{-\nu-i+k}
\frac{i!}{k!(i-2k)!}
(-1)^k (-2r)^{i-2k}.
\end{align*}
The signs simplify to one, yielding
\begin{equation}
\frac{\partial^i}{\partial r^i}(1-r^2)^{-\nu}
=
i!\sum_{k=0}^{\lfloor i/2\rfloor}
\frac{(\nu)_{i-k}}{k!\,(i-2k)!}\,
(2r)^{i-2k}\,
(1-r^2)^{-\nu-i+k}.
\end{equation}
It should be noted that $\phi_1^{(i,j)}$ rapidly grows very large ($n_{20} \sim \mathcal O(10^{17})$).

Turning to the $f_2^\alpha(r, \eta)$ and $f_3^\beta(r, \eta)$
components, we first compute the $\eta$ derivatives
\begin{align}
  \partial^j_\eta f_2^\alpha(r, \eta) = (\alpha)_{(j)} (2r)^j
  (f_2)^{\alpha-j} \\
  \partial^j_\eta f_3^\beta(r, \eta) = (\beta)_{(j)} (-2r)^j (f_3)^{\beta-j}.
\end{align}
Suppressing the explicit $(r, \eta)$ dependence on $f_2$ for
notational brevity, consider
\begin{align}
  \partial^i_r \partial_\eta^j f_2^\alpha(r, \eta) \eqqcolon \phi_2^{(i,j)}(r, \eta)&= (\alpha)_{(j)}
  \partial_r^i [(2r)^j f_2^{\alpha-j}] \cr
  &= (\alpha)_{(j)} 2^j \sum_{k=0}^{i} \binom{i}{k}[\partial_r^k r^j
  \cdot \partial_r^{i-k} f_2^{\alpha-j}]\cr
  &= (\alpha)_{(j)} 2^j \sum_{k=0}^{i} \binom{i}{k}[(j)_{(k)} r^{j-k}
  \cdot \partial_r^{i-k} f_2^{\alpha-j}]
  \label{f_2_deriv_liebniz}
\end{align}
where to obtain the second line from the first we use a special case
of \cref{liebniz_multivar}, namely
\begin{equation}
  \partial_r^n (f(r) g(r)) = \sum_{k=0}^n \binom{n}{k} \partial_r^k
  f(r) \partial_r^{n-k}g(r).
\end{equation}

The last non-trivial term in \cref{f_2_deriv_liebniz} is
$\partial_r^{i-k} f_2^{\alpha-j}$. Since the exponent is in general
not an integer, we cannot use the Leibniz rule, and thus must employ
the Fa\`a di Bruno formula, which states:
\begin{equation}
  \label{eqn:faa_di_bruno}
  \partial_r^a g_1(g_2(r)) = \sum_{s=1}^a g_1^{(s)}(g_2(r)) B_{a,
  s}(g_2^{(1)}(r), \ldots, g_2^{(a - s + 1)}(r)),
\end{equation}
where $B_{a,s}$ is an ordinary Bell polynomial. Letting $g_1(u) \coloneqq u^{\alpha-j}$ and $g_2 \coloneqq f_2(r, \eta)$, we may write
\begin{equation}
  \partial_r^{i-k} f_2^{\alpha - j}(r, \eta) = \sum_{s=1}^{i-k}
  ({\alpha - j})_{(s)} f_2^{{\alpha - j}-s}(r, \eta) B_{i-k,
  s}(f_2^{(1)}(r, \eta), \ldots, f_2^{(i-k - s + 1)}(r, \eta)).
\end{equation}
This simplifies further since there are only two non-zero
$r$-derivatives of $f_2(r, \eta)$, namely
\begin{equation}
\partial_r^{a} f_2(r, \eta) =
    \begin{cases}
    2r+2\eta &\quad \text{if } a = 1, \\
    2 &\quad \text{if } a = 2, \\
    0 &\quad \text{if } a \geq 3. \\
    \end{cases}
\end{equation}
Using the definition of the Bell polynomial, $\partial_r
f_2^{\alpha-j}$ may be written as
\begin{align}
  \partial_r^{i-k} f_2^{\alpha - j} &= \sum_{s=1}^{i-k} ({\alpha -
  j})_{(s)} f^{{\alpha - j}-s}_2(r, \eta) \sum_{\substack{j_1+j_2=s
  \\ j_1+2j_2=i-k}} \frac{(i-k)!}{j_1! j_2!} (f_2'(r, \eta))^{j_1}
  \bigg(\frac{f_2''(r, \eta)}{2!}\bigg)^{j_2}\\
  &= \sum_{s=\lceil \frac{i-k}{2} \rceil}^{i-k} (\alpha-j)_{(s)}
  f^{\alpha-j-s}_2 (r, \eta) \frac{(i-k)!}{(2s-i+k)! (i-k-s)!}
  (2r+2\eta)^{2s - i+k}\label{eqn:r_deriv_f2}\;,
\end{align}
where the second equality is realised by solving the simultaneous
equations in the sum index. Hence, one arrives at an expression for
$\phi_2^{(i,j)}(r, \eta)$, namely
\begin{align}
  \phi_2^{(i,j)}(r, \eta) &= (\alpha)_{(j)} \, 2^j\, \sum_{k=0}^i
  \bigg[ \binom{i}{k} (j)_{(k)} r^{j-k} \times \\ &\times \sum_{s=\lceil
    \frac{i-k}{2} \rceil}^{i-k} (\alpha-j)_{(s)} f^{\alpha-j-s}_2 (r,
  \eta) \frac{(i-k)!}{(2s-i+k)! (i-k-s)!} (2r+2\eta)^{2s - i+k}\bigg]. \nonumber
\end{align}

Similar reasoning can be applied to $f_3(r, \eta)$, resulting
in analogues of \cref{f_2_deriv_liebniz} and \cref{eqn:r_deriv_f2},
\begin{align}
  \partial^i_r \partial_\eta^j f_3(r, \eta)\eqqcolon \phi_3^{(i, j)}(r, \eta) &= (\beta)_{(j)}
  \partial_r^i [(-2r)^j f_3^{\beta-j}] \cr
  &= (\beta)_{(j)} (-2)^j \sum_{k=0}^{i} \binom{i}{k}[(j)_{(k)}
  r^{j-k} \cdot \partial_r^{i-k} f_3^{\beta-j}]
  \label{f_3_deriv_liebniz},
\end{align}
where
\begin{equation}
  \partial_r^{i-k} f_3^{\beta - j} = \sum_{s=\lceil \frac{i-k}{2}
  \rceil}^{i-k} (\beta-j)_{(s)} f^{\beta-j-s}_3 (r, \eta)
  \frac{(i-k)!}{(2s-i+k)! (i-k-s)!} (2r-2\eta)^{2s - i+k}.
\end{equation}
Combining the pieces,
\begin{multline}
  \phi_3^{(i,j)}(r, \eta) = (-2)^j (\beta)_{(j)}\sum_{k=0}^i
  \bigg[\binom{i}{k} (j)_{(k)} r^{j-k} \\ \sum_{s=\lceil
    \frac{i-k}{2} \rceil}^{i-k} (\beta-j)_{(s)} f^{\beta-j-s}_3 (r,
  \eta) \frac{(i-k)!}{(2s-i+k)! (i-k-s)!} (2r-2\eta)^{2s - i+k}\bigg].
\end{multline}

We now turn to the Gegenbauer polynomial $g(r)$. Notice that the
Gegenbauer polynomial is independent of $r$, meaning we only need to
consider the $m-i = 0$ contribution.
Using the identity
\begin{equation}
  \partial_\eta C_n^\lambda (\eta) = 2\lambda C_{n-1}^{\lambda+1} (\eta),
\end{equation}
we find by inspection
\begin{align}
  \partial_\eta^{n-j} g(r, \eta) &= \partial_\eta^{n-j} C_\ell^\nu(\eta) \cr
  &= 2^{n-j} (\nu)_{n-j}\,C^{\nu+n-j}_{\ell-n+j},
\end{align}
where to denote the rising factorial we have used  $(\nu)_{j} \coloneqq \nu
(\nu+1)\cdots(\nu+j-1)$ as in \cref{sec:deriv_approach}.

Putting everything together,
\begin{align}
  \label{eqn:partial_h_tilde}
  \partial_r^m \partial_\eta^n \tilde h^{\Delta_{ij}, \Delta_{kl}}_{\ell}
  &= \frac{(-1)^\ell \ell!}{(2\nu)_\ell} \partial_r^m \partial_\eta^n \big(f(r,
  \eta) g(r, \eta)) \cr
  &= \frac{(-1)^\ell \ell!}{(2\nu)_\ell} \sum_{j=0}^{n}
  \binom{n}{j} \bigg[\sum_{\substack{i_1+i_2+i_3=m\\j_1+j_2+j_3=j}}
    \frac{m!}{i_1!i_2!i_3!} \frac{j!}{j_1! j_2! j_3!}
    \phi_1^{(i_1,j_1)}\phi_2^{(i_2,j_2)} \phi_3^{(i_3,j_3)}
  \bigg] \times \nonumber\\ &\qquad\qquad\qquad\qquad \times [2^{n-j} (\nu)_{n-j}\,
  C^{\nu+n-j}_{\ell-n+j}].
\end{align}
In \texttt{GoBlocks}, \cref{eqn:final_recursive_rel} is implemented with
  $\partial_r^m\partial_\eta^n\tilde h^{\Delta_{ij}, \Delta_{kl}}_{\ell}$
defined in \cref{eqn:partial_h_tilde}.

\section{Derivatives of \texorpdfstring{$r, \eta$}{r, eta} with
Respect to \texorpdfstring{$z, \bar z$}{z, z-bar}}
\label{app:z_zb_r_eta_rosetta}

\texttt{GoBlocks} computes derivatives with respect to
the radial coordinates $r, \eta$. In order to use the results in the non-convex optimisation problem arising from the primal formulation (or compare with \texttt{scalar\textunderscore blocks}), one must
transform derivatives to $z, \bar z$ coordinates. In this appendix we derive the necessary transformation formulas.

Define
\begin{equation}
  \phi(z) \coloneqq \frac{z}{(1 + \sqrt{1-z})^2}
\end{equation}
and denote the radial coordinates as
\begin{equation}
  \eta(z, \bar z) \coloneqq \frac{\phi(z) + \phi(\bar z)}{2r(z, \bar z)}
  \;,\qquad r(z, \bar z) \coloneqq \sqrt{\phi(z) \phi(\bar z)}.
\end{equation}
In order to simplify calculations, define
\begin{align}
  \phi_1(z, \bar z) &\coloneqq \frac{\sqrt{z}(1+\sqrt{1-\bar z})}{\sqrt{\bar
  z} (1+\sqrt{1-z})}, \\ 
  \phi_2(z, \bar z) &\coloneqq \frac{\sqrt{z \bar
  z}}{(1+\sqrt{1-z})(1+\sqrt{1- \bar z})}, \\ 
  \phi_3(z, \bar z) &\coloneqq
  \sqrt{z} (1 + \sqrt{1 - \bar z}),
\end{align}
allowing one to write
\begin{equation}
  \eta(z, \bar z) = \frac{1}{2} (\phi_1(z, \bar z) + \phi_1(\bar z,
  z)), \qquad r(z, \bar z) = \phi_2(z, \bar z),
\end{equation}
and
\begin{equation}
  \phi_1 (z, \bar z) = \frac{\phi_3(z, \bar z)}{\phi_3(\bar z, z)}.
\end{equation}

Our aim is to calculate $  \partial_m \bar \partial_n
r(z, \bar z)$, $\partial_m \bar \partial_n \eta(z, \bar z)$, where
$\partial_m \coloneqq \frac{\partial^m}{\partial z^m}$, and $\bar
\partial_n \coloneqq \frac{\partial^n}{\partial \bar z^n}$. Focus first on the constituents of the $\eta$ derivatives
\begin{equation}
      \bar\partial_n \phi_3  = \sqrt{z}\bigl[\delta_{n,0} + \bar\partial_n\sqrt{1-\bar z}\bigr].
\end{equation}
Noticing that
\begin{equation}
  \bar \partial_n (\sqrt{1 - \bar z}) = (-1)^n (1/2)_{(n)} (1-\bar
  z)^{\frac{1}{2} - n}\;,
\end{equation}
the derivative becomes
\begin{equation}
  \bar \partial_n \phi_3(z, \bar z) = \sqrt{z} \big(\delta_{n, 0} +
  (-1)^n (1/2)_{(n)} (1 - \bar z)^{\frac{1}{2} - n}\big).
\end{equation}
Taking the derivative with respect to $z$, this becomes
\begin{equation}
  \partial_m \bar \partial_n \phi_3(z, \bar z) = (1/2)_{(m)}
  z^{\frac{1}{2} - m } \left(\delta_{n,0} + (-1)^n (1/2)_{(n)}
  (1-\bar z)^{\frac{1}{2} -n}\right).
\end{equation}

In order to obtain the full $\eta(z, \bar z)$ derivative, one must
consider the derivative $\partial_m \bar \partial_n
\left(\frac{\phi_3(z, \bar z)}{\phi_3(\bar z, z)}\right)$, where it
is convenient to utilise the Leibniz rule, namely
\begin{equation}
  \partial_m \bar \partial_n [f(z, \bar z) g(z, \bar z)] =
  \sum_{i=0}^m \sum_{j=0}^n \binom{m}{i} \binom{n}{j} \partial_i \bar
  \partial_j f(z, \bar z) \partial_{m-i} \bar \partial_{n-j} g(z, \bar z).
\end{equation}
In this case, one can identify $f(z, \bar z) \coloneqq \phi_3(z, \bar z)$ and
$g(z, \bar z) \coloneqq \phi^{-1}_3(\bar z, z)$. We turn our attention now to
$\partial_{m-i} \partial_{n-j} g(z, \bar z)$. In order to calculate
\begin{equation}
  \partial_a \bar \partial_b \phi_3^{-1} (\bar z, z),
\end{equation}
consider simply the $\bar z$ derivatives first
\begin{equation}
  \bar \partial_b \phi_3^{-1}(\bar z, z) = (1 + \sqrt{1-z})^{-1} \bar
  \partial_b \bar z^{-\frac{1}{2}} = (-1/2)_{(b)} (1 +
  \sqrt{1-z})^{-1} \bar z^{-(\frac{1}{2} + b)},
\end{equation}
before taking $z$ derivatives,
\begin{equation}
  (-1/2)_{(b)}\,\partial_a \left[(1 + \sqrt{1-z})^{-1}\right] \bar z^{-(\frac{1}{2} + b)}.
\end{equation}
The last derivative $\partial_a\left[(1+\sqrt{1-z})^{-1}\right]$ can be simplified by employing the Fa\`a di Bruno formula
\cref{eqn:faa_di_bruno}.
Let $g_1(u) \coloneqq (1+u)^{-1}$ and $g_2(z) \coloneqq \sqrt{1-z}$. The derivatives
of $g_1(u)$ and $g_2(z)$ are
\begin{equation}
  g_1^{(s)}(u) = (-1)^{s} s! (1+u)^{-(1+s)}\;, \qquad g_2^{(s)}(z)
  = (-1)^s (1/2)_{(s)} (1-z)^{\frac{1}{2} - s}.
\end{equation}
The final formula is thus
\begin{equation}
  \psi_{a,b}(\bar z, z) \coloneqq \partial_a \bar \partial_b \phi_3^{-1} (\bar z, z) = \tilde
  \psi_{a}(z) \bigg[\bar z^{-(\frac{1}{2} + b)}(-1/2)_{(b)}\bigg],
\end{equation}
where $\tilde \psi_{a}(z)$ is defined such that
\begin{align}
  \tilde \psi_{0}(z) & \coloneqq  (1+\sqrt{1-z})^{-1},\cr
  \tilde \psi_{a\ge 1}(z) & \coloneqq \sum_{s=1}^a (-1)^s s!
  (1+g_2(z))^{-(1+s)} B_{a, s}(g_2^{(1)}(z), \ldots, g_2^{(a - s + 1)}(z)),
\end{align}
and  $B_{a,s}(x)$ are generalised exponential Bell polynomials.

Combining results,
\begin{align}
  \partial_m \bar \partial_n \phi_1(z, \bar z) = \sum_{i=0}^m
  \sum_{j=0}^n \binom{m}{i} \binom{n}{j} (1/2)_{(i)} \;
  z^{\frac{1}{2} - i } \left(\delta_{j,0} + (-1)^j (1/2)_{(j)}
  (1-\bar z)^{\frac{1}{2} -j}\right)
  \psi_{m-i,n-j}(\bar z, z)
\end{align}
and  one can realise a formula for the tower of $\eta$ derivatives by noting that
\begin{equation}
  \partial_m \bar \partial_n \eta(z, \bar z) = \frac{1}{2}
  (\partial_m \bar \partial_n \phi_1(z, \bar z) + \partial_m \bar
  \partial_n \phi_1(\bar z, z)).
\end{equation}

We can now turn to the $r$
derivatives. One begins by writing
\begin{equation}
  r(z, \bar z) = f_1(z, \bar z) f_2(z) f_3(\bar z),
\end{equation}
where
\begin{align}
  f_1(z, \bar z) &\coloneqq (z \bar z)^\frac{1}{2}\\
  f_2(z) &\coloneqq (1 + \sqrt{1-z})^{-1} \\
  f_3(\bar z) &\coloneqq (1+\sqrt{1-\bar z})^{-1}.
\end{align}
Using the Leibniz formula once more,
\begin{equation}
  \partial_m \bar \partial_n r(z, \bar z) =
  \sum_{\substack{m_1+m_2+m_3=m \\ n_1+n_2+n_3=n}} \frac{m!}{m_1!
  m_2! m_3!} \frac{n!}{n_1! n_2! n_3!} (\partial_{m_1} \bar
  \partial_{n_1} f_1(z, \bar z)) (\partial_{m_2} \bar \partial_{n_2}
  f_2(z)) (\partial_{m_3} \bar \partial_{n_3} f_3(\bar z)).
\end{equation}
Consider first $\partial_{m_1} \bar \partial_{n_1} f_1(z, \bar z)$:
\begin{align}
\label{eqn:f1_deriv}
\partial_{m_1} \bar \partial_{n_1} f_1(z, \bar z) &=
  \partial_{m_1} \bar \partial_{n_1} (z \bar z)^\frac{1}{2}\cr
    &= \sum_{m=0}^{m_1} \sum_{n=0}^{n_1} \binom{m_1}{m}
    \binom{n_1}{n} \partial_m \bar \partial_n (z^\frac{1}{2})
    \partial_{m_1-m} \bar \partial_{n_1-n} (\bar z^\frac{1}{2}).
  \end{align}
  The innermost double derivatives of the $z, \bar z$ terms may be realised as
  \begin{align}
    \partial_m \bar \partial_n (z^\frac{1}{2}) &= \delta_{n,0}
    (1/2)_{(m)} z^{\frac{1}{2} - m}\\
    \partial_{m_1-m} \bar \partial_{n_1-n}(\bar z^\frac{1}{2}) &=
    \delta_{m_1 - m, 0}\,(1/2)_{(n_1-n)}\, \bar z^{\frac{1}{2} - n_1 + n}.
  \end{align}
  Combining with \cref{eqn:f1_deriv},
  \begin{equation}
    \partial_{m_1} \bar \partial_{n_1} (z \bar z)^\frac{1}{2} =
    (1/2)_{(m_1)} (1/2)_{(n_1)} z^{\frac{1}{2} - m_1} \bar
    z^{\frac{1}{2} - n_1}.
  \end{equation}

Then, focussing our attention on the $f_2(z)$ and $f_3(\bar z)$
derivatives, we notice that
\begin{equation}
\partial_{m_2} \bar \partial_{n_2} f_2(z) = \delta_{n_2, 0}
\partial_{m_2} [(1+\sqrt{1-z})^{-1}] = \delta_{n_2, 0} \tilde
\psi_{m_2}(z)
\end{equation}
and
\begin{equation}
\partial_{m_3} \bar \partial_{n_3} f_3(\bar z) = \delta_{m_3,
0} \partial_{n_3} [(1+\sqrt{1-\bar z})^{-1}] = \delta_{m_3,
0} \tilde \psi_{n_3}(\bar z).
\end{equation}

The final formula for the $r$ derivatives is then
\begin{equation}
\partial_m \bar \partial_n r(z, \bar z) = \sum_{\substack{m_1
+ m_2 = m \\ n_1 + n_2 = n}} \frac{m! n!}{m_1! m_2! n_1!
n_2!} (1/2)_{(m_1)} (1/2)_{(n_1)} z^{\frac{1}{2} - m_1} \bar
z^{\frac{1}{2} - n_1} \tilde \psi_{m_2}(z) \tilde \psi_{n_2}(\bar z).
\end{equation}

\begin{figure}[t]
    \centering
    \includegraphics[width=0.75\linewidth]{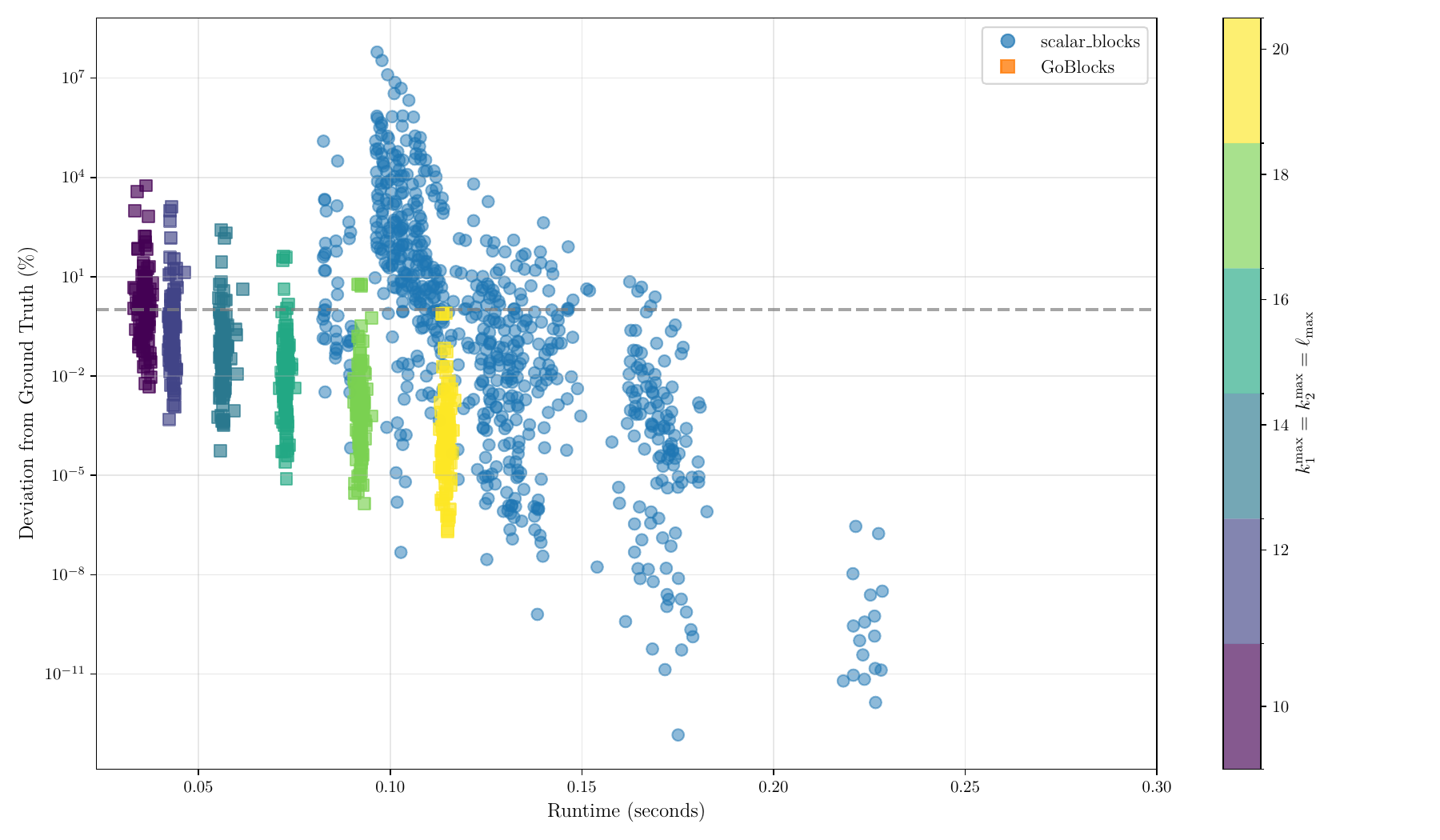}
    \caption{Runtime versus accuracy for \sbl and \texttt{GoBlocks} with a selection of hyperparameters, with the 1\% accuracy level depicted by a dashed horizontal line.}
    \label{fig:tuning_ktriple}
\end{figure}

\section{Tuning \gbl for Optimal Performance}
\label{app:tuning_ktriple}
As discussed in \cref{sec:timing}, the performance of \gbl is predominantly controlled by $k_{1,2}^\text{max}$ and $\ell_\text{max}$. In the absence of a canonical strategy for configuring these parameters, it is useful to tune them in order to minimise the runtime for a desired level of block accuracy. This is particularly relevant when calling \gbl many times, as e.g. in the process of solving the crossing equations via stochastic optimisation. To some tolerance, there exists an equivalence class of combinations of $k_{1,2}^\text{max}, \ell_\text{max}$ resulting in a given level of performance. A well-motivated technique to find the optimal parameter combination is to perform a Bayesian sweep in the space of configurations. Alternatively, one may adopt a na\"ive approach and set each of the aforementioned parameter to a common value. \cref{fig:tuning_ktriple} presents block accuracies for various parameters compared to those evaluated with \texttt{scalar\_blocks} at varying orders, for common values of $k_{1,2}^\text{max}$ and $\ell_\text{max}$. We stress that while the hyperparameter combinations depicted in \cref{fig:tuning_ktriple} are likely sub-optimal, for a band of accuracies, there exist parameter combinations where \gbl is more computationally efficient than \texttt{scalar\_blocks}. Despite this, for large $n_\text{max}$ and high block accuracies, \sbl offers better performance.

\end{appendix}

\bibliographystyle{JHEP}
\bibliography{bib}

\end{document}